\documentclass[pre,aps,twocolumn,showpacs,floatfix]{revtex4-2}

\usepackage{amsmath}
\usepackage{amssymb}
\usepackage{bm}
\usepackage{graphicx,subfigure}
\usepackage{epsf,epsfig}
\usepackage{hyperref}
\usepackage{color}
\usepackage{enumerate}

\newcommand{\bq}{\begin{equation}}
\newcommand{\eq}{\end{equation}}
\newcommand{\bqn}{\begin{eqnarray}}
\newcommand{\eqn}{\end{eqnarray}}
\newcommand{\nb}{\nonumber}
\newcommand{\lb}{\label}

\begin{document}

\title{Large-scale cortex-core structure formation in brain organoids}
\author{Ahmad Borzou$^1$}
\author{J. M. Schwarz$^{1,2}$}
\affiliation{$^1$ Department of Physics and BioInspired Institute,  Syracuse University, Syracuse, NY USA,
$^2$ Indian
Creek Farm, Ithaca, NY, USA}
\date{\today}
\begin{abstract}
Brain organoids recapitulate a number of brain properties, including neuronal
diversity. However, do they recapitulate
brain structure? Using a hydrodynamic description for cell nuclei
as particles interacting initially via an effective, attractive force
as mediated by the
respective, surrounding cytoskeletons, we quantify structure development in brain
organoids to determine what physical mechanism regulates the number of cortex-core structures.
Regions of cell nuclei overdensity in the linear regime drive the initial
seeding for cortex-core structures, which ultimately develop in the non-linear
regime, as inferred by the emergent form of an effective
interaction between cell nuclei and with the extracellular
environment, as mediated by a dynamic cytoskeleton. Individual cortex-core structures then provide a basis upon
which we build an extended version of the buckling without
bending morphogenesis (BWBM) model, with its proliferating cortex and
constraining core, to predict foliations/folds of the
cortex in the presence of a nonlinearity due to cortical cells actively
regulating strain. In doing so, we obtain asymmetric foliations/folds with respect to the
trough (sulci) and the crest (gyri). In addition to laying new groundwork for the design
of more familiar and less familiar brain structures,  the hydrodynamic
description for cell nuclei during the initial stages of brain
organoid development provides an intriguing
quantitative connection with large-scale structure formation in the universe.   
\end{abstract}                                                                                   
\maketitle

\section{Introduction} 
What physical mechanisms are at play in determining
  brain structure? In humans, the beginning of brain structure begins
  about two weeks after fertilization with the formation of a neural
  plate ~\cite{DevNervousSystemBook}. The neural plate then folds
  inward on itself to form a neural tube.  From this neural tube,
  different brain regions, such as the forebrain and the hindbrain
  emerge. In the forebrain, the proliferating progenitor cells in the
  innermost part of the tube form the ventricular zone, with extended,
  radial glial cells linking the cells in the ventricular zone to the
  outer edge of the neural tube.  It is the extended radial glial
  cells that the inner progenitor cells crawl along to reach the outer
  part of the forebrain.  As they do so, they differentiate to become
  neurons and form a cortex, consisting of six layers of cells, around
  20 weeks later. In the hindbrain,  the proliferation of progenitor
  cells occurs in the outer region with their migration towards the
  center of the structure. In humans, both the cerebral cortex (from
  the forebrain) and the cerebellar cortex (from the hindbrain)
  undergo shape changes in the form of folds or foliations. 

Until recently, many of the biophysical models for brain structure have focused on the later stages of brain
shape development, namely the development of folds of the cerebrum and
the cerebellum~\cite{Richman1975,VanEssen1997,Bayly2013,Manyuhina2014,Budday2015,Mota2015,Tallinen2016,Lejeune2016,Lejeune2019,Engstrom2018,Lawton2019}. 
These models essentially divide into two camps. The first camp consists of nonlinear elastic models with
differential swelling mimicking cell growth and generating compressive
forces~\cite{Richman1975,Bayly2013,Budday2015,Tallinen2016,Lejeune2016,Lejeune2019}. While the
second camp focuses on tension-based, multi-phase models in the
presence of cell growth and 
generating tensile forces~\cite{VanEssen1997,Manyuhina2014,Engstrom2018,Lawton2019}. Experiments on ferret brains, whose folds
develop {\it ex utero}, appear to rule out the
initial version of the tension-based models by observing the displacement of
brain tissue in response to cuts in particular directions~\cite{Xu2010}. However, there exists a
revised version of the initial tension-based model with a different
direction of the tension that has yet to be tested experimentally~\cite{Manyuhina2014}.  Moreover,
experimental studies on the developing mouse cerebellum, or the little
brain, validate a new
tension-based model, dubbed ``buckling without bending'', and rule out
a nonlinear elastic model with differential
swelling~\cite{Engstrom2018,Lawton2019}.

\begin{figure}[t]
\centering
\includegraphics[width=\columnwidth]{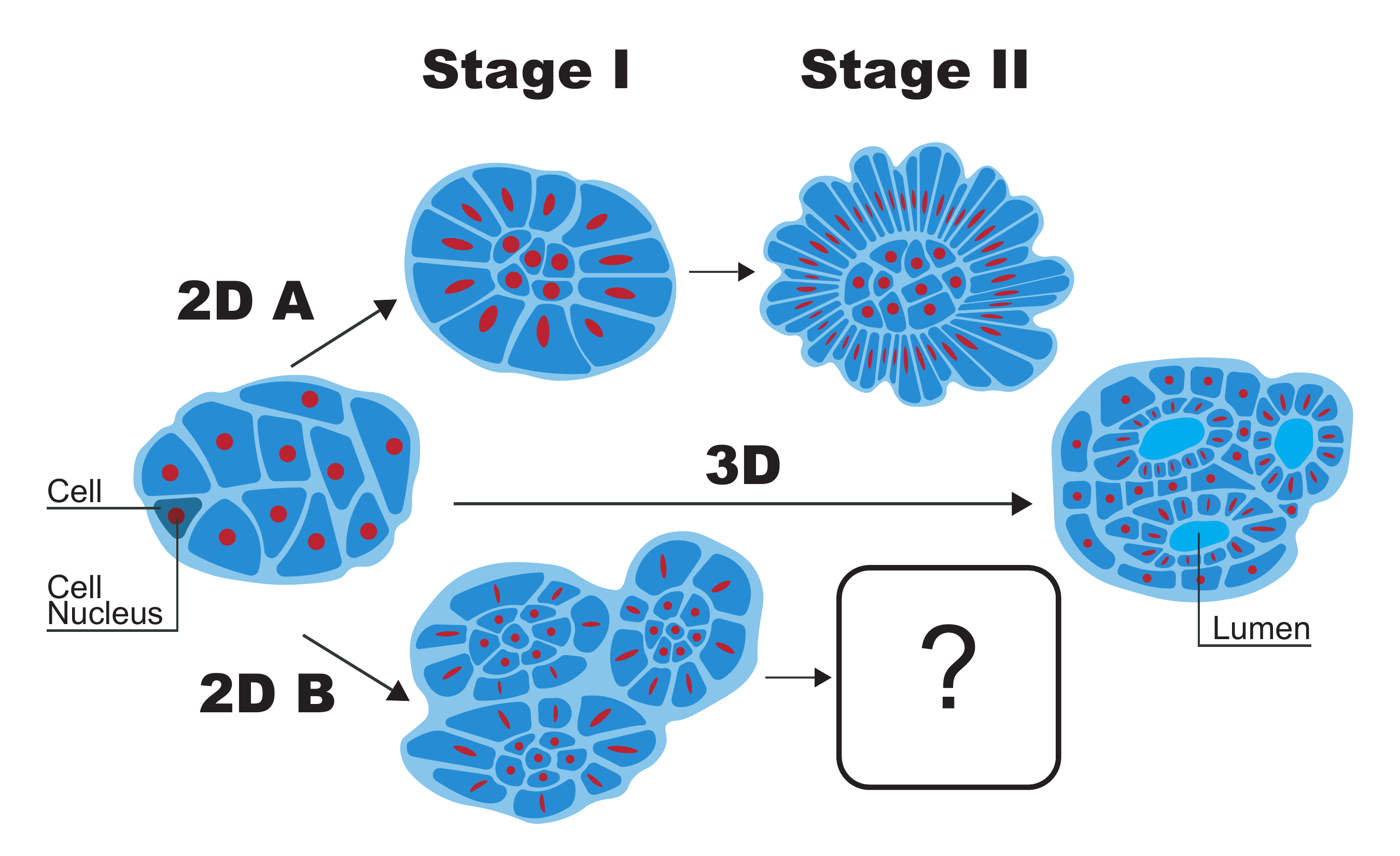}
\caption{A simplified version of different possible organoid shapes in
  quasi-2D confinement (2DA and
2DB) and not in quasi-2D confinement (3D). The cortex-core structure
consists of globular-shaped cells surrounded by cells extended
radially outward. One such structure emerges in the 2D A pathway
during Stage I. Credit: Savana Swoger.}
\end{figure}

While many biophysical studies of brain structure focus on the folds of the cerebrum
or the cerebellum, the emergence of the base
structure of a cortex-core at earlier stages of development is also
very important. 
It is this emergence that we will focus on for the first
part of the manuscript using a minimal model. Interestingly, a new {\it in vitro} brain system, namely brain
organoids~\cite{Lancaster2013}, provides a testing ground for
understanding the emergence of brain
structure~\cite{Lancaster2021,Karzbrun2018}, as well as recapitulate such phenomena as
neuronal diversity~\cite{Velasco2019} and neuronal firing
patterns~\cite{Trujillo2019}. 
One such study focuses on quasi-two-dimensional brain organoids that start
as a relatively isotropic clump of cells, which is then inserted between
two ``plates'' separated by approximately 150 $\mu$m~\cite{Karzbrun2018}. Within several days, the clump morphs into several
sub-structures each with a core with globular, polyhedral-shaped cells and radially-stretched
cells surrounding each core to form a cortex, or, there is only one
cortex-core structure.  See Fig. 1 in which latter result is labelled
pathway ``2D A'' and the former, ``2D B''. After this initial shape
change in the ``2D A'' pathway and the cortex-core structure continues
to grow, a
foliation amongst the radially-stretched cells emerges. It is for this
latter foliation stage that the researchers develop a theory, which is
rooted in elasticity theory~\cite{Karzbrun2018}. 

Here, we will first take a step back from cortex foliation and address how is it that at times
multiple cortex-core structures emerge, while in other cases, just one
cortex-core structure emerges.  Answering this question will give us
insight into cortex-core formation. 
Then, we will characterize the foliation
of the cortex with the ``buckling without bending'' morphogenesis
(BWBM) model since it is already able to quantitatively capture cortical foliation in blebbistatin-treated
brain organoids~\cite{Engstrom2018}.  One, therefore, wonders whether or not such
an approach is applicable in the untreated case. 

To explore how brain
organoids acquire their cortex-core 
structure, which we dub Stage I, {\it and} subsequent cortex
foliation, or Stage
II (see Fig. 1), we build continuum models for each respective
stage. We do so since once the cortex-core structure forms, that
structure provides the basis for coarse-graining at a different scale
to arrive at an extension of the BWBM model. As for Stage I, many cellular
aggregates often demonstrate fluid-like behavior \cite{Beaune8055,stirbat2013fine}, though
viscoelasticity is also observed \cite{PhysRevLett.104.218101}. Given the cellular fluid-like
behavior, an appropriate model
falls under the hydrodynamics domain.  At this point, it is tempting
to go to a more detailed cellular-based model. However, we will take a less-detailed
approach since it is not entirely clear if we can address the one
versus many cortex-core structures within a detailed computational model at the
outset,  given the limitation of finite size. 

Therefore, we seek a more minimal
approach in which cell nuclei are particles. The cell nuclei interact
with other cell nuclei {\it indirectly} via the surrounding active cytoplasm of,
say, two cells, interacting with each other via their cortical
tension, which ultimately drives the cell-cell interaction. Cell nuclei also
interact indirectly with their 
extracellular surroundings. In other
words, there is an {\it effective, active force} on cell nuclei due to
cell-cell interactions an {\it effective, active force} on cell
nuclei due to the extracellular environment. 
Specifically,  since cell nuclei are observed to move toward each other during what is called the linear regime, we will assume that there is an initial, effective, short-range attraction
between cell nuclei.
As the organoid evolves, given the dynamic and mechanosensitive nature of the active,
cytoskeleton generating forces, we use several
observations from experiments to determine the emergent form of
the effective, active force on cell nuclei due to cell-cell and
cell-environment interactions during the non-linear regime.  From the emergent form of the effective, active force
we will infer cortex-core structures. We will use hydrodynamic equations to describe both
regimes, which will correspond to a linear regime and a non-linear
regime.

Intriguingly, this vantage point draws parallels with cosmological models of large-scale structure formation in our
universe at the quantitative level.  Both have the same dynamical equation for the single-particle
distribution function.  Moreover, the initial and final states of both
systems are very similar; both systems start from an initially uniform
number density of particles and end in spherical shape structures.
This is despite the differences such as different types of interactions, as well as
the presence of growth, dissipation, and noise, for example, for
living matter.

Once cortex-core structures emerge,
  such individual structures provide the basis for the BWBM
  model in which cellular growth plays a key role. The BWBM model
  consists of an incompressible core with a growing cortex and several
  other mechanical aspects of the brain organoid structure, namely,
  the mechanics of the surrounding Matrigel and cortical cells under
  tension. An initial, linear version of the BWBM model provided a
  physical basis for the unusual cortical thickness variations in
  blebbistatin-treated brain organoids---variations that cannot be
  readily explained by a purely elastic model. In addition, the
  foliations were more scalloped, or more asymmetric, in the
  blebbistatin-treated case as compared to the untreated case, and so
  a linear BWBM model, with its symmetric foliations was
  reasonable. Indeed, blebbistatin inhibits myosin-II, and so the
  intra-cellular tension in cells making up both the cortex and the
  core decreases~\cite{Karzbrun2018}. At higher tensions, nonlinearities in tension are
  more likely to become relevant, which is what we explore
  here. For context, recent work has been done to take into account
  nonlinearity in tension in radial glial cells in the developing
  cerebellum~\cite{Gandikota2021}, which are not present in the confined brain organoids at
  relevant time scales. We will explore a different form of 
  nonlinearity here. 

This paper is organized as follows.  In Section II, we uncover a
physical mechanism for cortex-core structure formation (Stage I). In Section III, we
quantify how foliations/folds in the cortex emerge.  We conclude with 
implications of our findings in Section IV.

\section{Stage I Model} 

Assuming that the brain organoid begins as an aggregate of fluid-like
cells, let us begin with the dynamical equation for the one-body distribution
function for cell nuclei given by 
\bqn
\lb{Eq:StochasticDynamicEq}
\frac{df}{dt} = \frac{\partial f}{\partial t} + \frac{\partial f}{\partial x^{i}}v^{i} + \frac{\partial f}{\partial v^{i}}\ddot{x}^{i} = C,
\eqn
where $C$ incorporates dissipation, cell
division, and fluctuations~\cite{Borzou2021}.  We treat the interaction
between cell nuclei as mediated by the surrounding cytoskeletons associated with two attached cells, for instance, as an
effective force on a cell nucleus. A surrounding cytoskeleton interacting
with the extracellular environment also leads to an effective force
on cell nuclei. In other words, even though nuclei do not interact
directly, they interact indirectly via their respective, surrounding cytoskeletons, so using
Newton's second law, the sum of the forces on the nuclei are such that
the net force on them may be nonzero should they go from not moving to
moving, for example. 

As for how the effective force between cell nuclei is generated, a
significant player is the contractile nature
of the actomyosin cortex with cells fusing when in a fluid phase
~\cite{Martin1975,Murrell2015}.  This phenomenon points to an
effective, attractive interaction between cell nuclei that is short range in
the sense that the interaction involves cells in contact. There are also effective interactions between the cell
nuclei and their passive environment as mediated by the cell
cytoskeleton~\cite{Humphrey2014}.  Therefore, we approximate both
cell-cell and cell-environment interactions as an effective force such that $\ddot{x}^{i}$ in
Eq.~\eqref{Eq:StochasticDynamicEq}  refers to the effective force on
cell nuclei as mediated by the cell cytoskeleton accounting for
factors such as 
actomyosin contractility and the surrounding environment. 

With this construction, there is a quantitative link with cosmology. In cosmology, to
study large-scale structure formation,  one constructs a dynamical equation for the time
evolution of matter as encoded by the one-body distribution function,
$f$, in the six-dimensional phase space of positions and velocities. See, for example, the Boltzmann equation~\cite{steinhardt2005physical} and the Jeans equation~\cite{binney2008galactic}.
In the Jeans equation in cosmology, $\ddot{x}^{i} = g^{i}$, where
$g^{i}$ denotes gravitational acceleration~\cite{binney2008galactic}.

Inspired by this mathematical link to cosmology, we split Stage I into linear and
non-linear eras.  Since cell nuclei densities are low initially,
Eq.~\eqref{Eq:StochasticDynamicEq} is linear.  We will show that the
magnitude and the form of the effective force do not change the
resulting structures' shape.  As soon as cell nuclei densities
increase,  the equation becomes non-linear and the exact form of the
effective force becomes relevant.  We will assume that the form of the
effective force can change with time as it is generated by a dynamic
cytoskeleton. Given certain observations from experiments in the
non-linear era, we will extract the effective
force from Eq.~\eqref{Eq:StochasticDynamicEq}.  The spatial
patterning of cell nuclei then emerges 
from the combination of (i) a robust evolution equation and (ii)
observation, and is rooted in the effective force that we derive.
Specifically, we show that at the beginning of the non-linear era, the
effective force between cell nuclei is attractive.  However, toward
the end of Stage I,  the effective force changes in nature and becomes
almost neutral at the center and repulsive beyond some characteristic radius of the spherical structures.

Let us now work towards a solution for Eq. (1). Since cell nuclei are roughly round shape in Stage I of brain organoid
formation,  we neglect their inherent structure and 
assumed that phase-space consists of positions and velocities only.  
Solving Eq.~\eqref{Eq:StochasticDynamicEq} analytically can be
challenging.  The more conventional approach is to solve its first two
moments of velocities leading to two differential equations coupling
the 
number density $\rho \equiv \int dv \, f$ , the bulk velocity
$\bar{v}^j \equiv \frac{1}{\rho}\int dv \, v^j\, f$,  and $\overline{vv}^{ij} \equiv \frac{1}{\rho}\int dv \, v^iv^j\, f$. To
proceed further, one can write a third differential equation for $\overline{vv}^{ij}$, which
depends on higher moments of $f$. Instead, we apply a data-driven
hydrodynamics approach to find a relationship between $\rho$ and
pressure for the cell nuclei~\cite{Borzou2021}.  To do so,  we define the stress tensor as 
$\sigma^{2ij}  \equiv \overline{vv}^{ij} - \bar{v}^i\bar{v}^j$ and
assume it is isotropic during the initial stages such that $\sigma^{2ij} = \sigma^2 \delta^{ij}$. From the observations
reported in Ref.\cite{Karzbrun2018}, we find that the pressure of the
cell nuclei
linearly depends on the number density with a proportionality
coefficient of $\sigma^2=0.1$ (see Appendix A).  
Therefore, the final form of the evolution equation set reads
\bqn 
\label{Eq:MomentEqs}
&&\partial_t \rho + \partial_i \left( \rho \bar{v}_i \right) = C_0,\nb\\
&&\partial_t \bar{v}_j + \sigma^2 \partial_j \rho + \bar{v}_i \partial_i \bar{v}_j + g_j = \frac{1}{\rho} \left(C_j - \bar{v}_j C_0 \right),
\eqn
where
$C_0 \equiv \int dv \, C$ accounts for cell division and the
noise,  and $C_j \equiv  \int dv \, C \, v_j$ accounts for
dissipation and noise. Since the number density experiences minimal
growth in the first 3 days of the experiment until the cortex-core 
structures are first observed~\cite{Karzbrun2018}, we assume $C_0$ only accounts for the noise. We also assume that the
noise terms obey 
\bqn
\langle C_{0_{\text{noise}}} \rangle = \langle
C_{j_{\text{noise}}} \rangle =0,
\eqn 
with 
\bqn
&&\langle C_{0_{\text{noise}}}(t,\vec{x}) C_{0_{\text{noise}}}(t',\vec{x}') \rangle
=\theta \delta(t-t')\delta^3(\vec{x}-\vec{x}'), \nb\\
&&\langle C_{i_{\text{noise}}}(t,\vec{x}) C_{j_{\text{noise}}}(t',\vec{x}') \rangle
=\gamma\delta_{ij}\delta(t-t')\delta^3(\vec{x}-\vec{x}'),
\eqn 
where $\theta$ and $\gamma$ determine the strength of each type of noise. For
effects of dissipation, see Appendix A.

Given the coupled,  non-linear equations above, we will divide the
brain organoid evolution into linear and non-linear regimes by solving
the linearized form to find the initial conditions for the non-linear
evolution.  However, in the non-linear regime, instead of deriving density growth
in terms of the forces, we use the exact form of the differential
equations,  and the observations,  to derive the evolution of
the effective forces on the cell nuclei as they may change over time,
unlike in cosmology.

\subsection{Stage I linear regime}   
We assume that the number density is
initially homogeneous with some small fluctuations, or  $\rho \equiv \rho_0 + \delta \rho$,
with $\delta \rho \ll \rho_0$.  Inserting this ansatz into
Eq.~\eqref{Eq:MomentEqs}, neglecting higher order terms, and
Fourier transforming, we find 
\bqn
\lb{Eq:delta_rho_t_x}
\delta \rho(t,\vec{x})=\int d^3k\,
e^{i\vec{k}\cdot\vec{x}} \, \tilde{\delta}(t,\vec{k}),
\eqn 
with 
\bqn
\tilde{\delta}(t,\vec{k}) = \tilde{\delta}(t=0,\vec{k})
\cosh\left(\sqrt{\rho_0 ( {\cal{L}}^{-1} -\sigma^2) } ~k t\right),
\eqn
where the Fourier transform of  ${g}_j$ is assumed to have the following general form  $\tilde{g}_j=- i k_j {\cal{L}}^{-1} \tilde{\delta}$.  
In a conservative system with long-range gravitational forces,  ${\cal{L}}^{-1} =
k^{-2}$.  For brain organoids, the forces are short-ranged and
non-conservative, so ${\cal{L}}^{-1}$ takes a more complex
form.  Nevertheless,  the result is not sensitive to the detailed form of the attractive force in the linear regime since only the first term of its Taylor expansion contributes to the results. 
We assume two generic forms below.

\begin{figure*}
\centering
\includegraphics[width=0.99\textwidth]{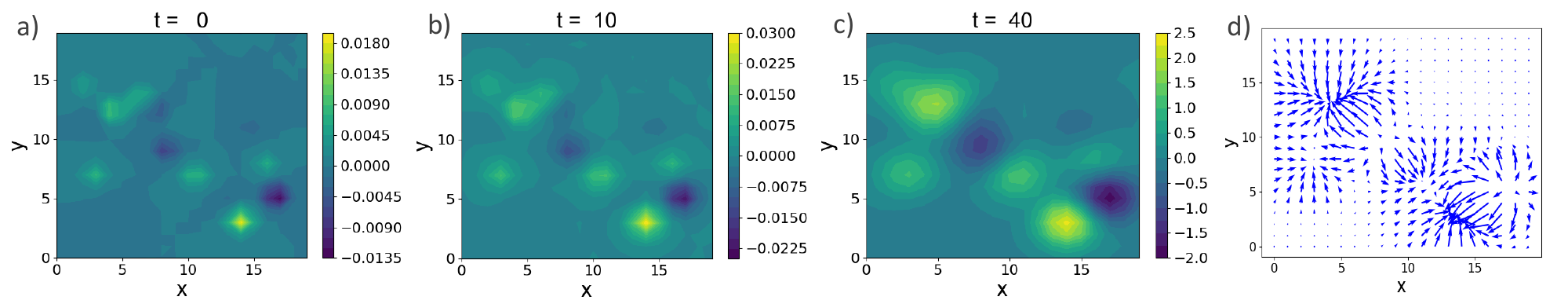}
\caption{a-c): The time evolution of $\delta\rho$ for
  ${\cal{L}}^{-1} = \frac{1}{k^2+0.1^2}$ with $\rho_0=1$,  $a=0.1$,
  and $b=0.1$. d): The corresponding force field at $t=40$. }
\lb{Fig:EvolutionOfRho}
\end{figure*}

To model the very initial overdensities,  we assume that at $t=0$ there exist $N$ point-like random fluctuations in the density such that 
\bqn
\delta\rho(t=0,\vec{x}) = \sum_{i=1}^N c_i
e^{-|\vec{x}-\vec{r}_i|^2},
\eqn 
where $\vec{r}_i$ and $c_i$ are random and denoting the location and magnitude of each density fluctuation.  All magnitudes satisfy $|c_i| \ll \rho_0$.  Therefore,  
\bqn
\tilde{\delta}(t=0,\vec{k})= \pi^{\frac{3}{2}}\sum_{i=1}^N c_i \,
e^{-i\vec{k}\cdot \vec{r}_i}\,e^{-k^2/4}.
\eqn 
Inserting all terms back into Eq.~\eqref{Eq:delta_rho_t_x},  the final solution for the time-evolved number density in the linear regime reads
\bqn
\lb{Eq:deltaRhoSolFinal}
\delta \rho (t,\vec{x}) &=& \frac{1}{2\pi^{\frac{1}{2}}} \sum_{i=1}^N \frac{c_i}{|\vec{x}-\vec{r}_i|} \int_0^{\infty} 
k \sin(k\, |\vec{x}-\vec{r}_i|) e^{-k^2/4}\nb\\
&&
\times \cosh \left(\sqrt{\rho_0\left({\cal{L}}^{-1}  - \sigma^2 \right)} ~k t
   \right)dk.
\eqn
While we do not know the exact form of the initial,
  effective force between cell nuclei, from observation, it is
  effectively short-range and contractile. So we write ${\cal{L}}^{-1}$ in the following general form
\bqn
\lb{Eq:Linv}
{\cal{L}}^{-1} = \sum_{n=1}^{\infty}\frac{l_n}{k^{2n}+b_n^{2n}},
\eqn
where $b_n^{-1}$ is an effective distance beyond which the force is zero.
While the final over/under-densities
slightly depend on the significant terms in ${\cal{L}}^{-1}$,   as far as the effective interactions are attractive, the
structures grow with a rather similar form.  

In the following,  we initially work with the first term of the sum in
Eq.~\eqref{Eq:Linv} and derive the final densities. Later, we repeat
the same calculation with an additional term in the sum to look for
sensitivity in the form of the effective force.  
After retaining the first term in the expansion of
${\cal{L}}^{-1}$,  we 
insert the randomly generated $c_i$ and $\vec{r}_i$ and integral numerically for every $\vec{x}$ over a finite area in the
$x-y$ plane to determine how $\delta \rho(x,y)$ changes in time.  We choose $\rho_0=1$,  $a=0.1$,  and $b=0.1$ to carry
out the calculation.  
The density evolution for
different time points is plotted in Fig.~\ref{Fig:EvolutionOfRho}.  Small
over-density  and under-density regions grow under the contractile forces of the cells
and create cortex-core seeds for the non-linear regime.
We also have repeated the calculations for ${\cal{L}}^{-1}$ containing the first {\it two} terms in
 eq.~\eqref{Eq:Linv}.  
The final density is presented in  Fig.~\ref{Fig:EvolutionOfRho_with2nd},  indicating that,  as long as the effective force between nuclei is attractive, the
larger-scale density structures grow with a rather similar form.  The
difference is more in the timing of the growth. The stronger the
force, the faster the structures form. The reason for the similar
spatial structure is that regardless of the exact form of the force, in the linear era, one can always perform a Taylor expansion and neglect the higher order terms. Hence, the effective force always enters the equations with the same form regardless of its exact form.

\subsection{Stage I non-linear regime}
At the end of the linear era, the
cell nuclei around each existing over-dense region 
start to migrate toward a center. However, unlike in the linear regime, results may indeed
  depend on the details of the net, effective force on cell
  nuclei. Given the dynamic nature of the cytoskeleton mediating 
  the effective force, combined with the existence of experimental
  data, we adjust 
  our approach and use a data-driven approach to derive the effective
  force on nuclei using our knowledge of density evolution from
  observations.  Our prediction for the emergent, effective force can
  be tested with additional experiments.

We now focus on
one of the over-dense centers and assume a spherical symmetric structure with  
$\rho \simeq \rho_0$, $\partial_t \rho \simeq 0$, $\bar{v}_r = - v_0$
and reset time to $t=0$. 
The evolution equations now become 
\bqn
\lb{Eq:MomentEqs2}
&&\frac{\partial \rho}{\partial t}+ \frac{2}{r}\rho\bar{v}_r + \partial_r \left( \rho \bar{v}_r \right) = C_0,\nb\\
&&\frac{\partial}{\partial t}  \bar{v}_r + \sigma^2 \partial_r \rho +\bar{v}_r \partial_r\bar{v}_r
+  g_r =\frac{1}{\rho} C_r ,
\eqn
where we have used the isotropic assumption to infer that in the
spherical coordinate system $\vec{\bar{v}}=(\bar{v}_r, 0)$.

We use
the first equation to solve for the bulk velocity in terms of the
number density and then make the following assumptions for the
final state of what becomes the cortex-core structure: $\rho
\rightarrow F(r)$,  
where $F(r)$ is the form of the number density observed around Day 3
of the experiment (see Fig.~\ref{Fig:Fr} ), and $\bar{v}_r
\rightarrow  0$.  Given the initial and final conditions, we construct
an analytic form for $\rho(t,r)$ so that we can ultimately determine
the interaction between the cell nuclei.  Unlike in the linear regime (and
in cosmology), the interactions between cell nuclei can change to the
cytoskeletal restructuring in response to interactions with other
cells and/or with the environment. Hence,  $g_r(t,r)$ is unknown as are
the damping effects in $C_r(t,r)$, 

With these assumptions, the time evolution for $\bar{v}_r(t,r)$ and
$g_r(t,r)-C_r(t,r)/\rho(t,r)$ can be determined. See the appendix for
details. The results for the latter are shown in Fig.~\ref{Fig:Forces}. 
The bulk velocity is initially position-independent and toward the
center.  Over time, it becomes a position-dependent function and
evolves toward zero.  The net force is toward the center of the core
initially, but changes nature over time and becomes position-
dependent.  By Day 3, near the edge, the net
force is outward, indicating that the nuclei are being indirectly pulled on by
the extracellular environment,  i.e. the cell cytoskeleton
has developed subcellular structures to attach to the extracellular environment. We have
not assumed the existence of such an effect 
but derived it based on data and the
theoretical framework.  Experiments can measure this net force via
laser ablation. 

Subtracting from the net force the assumed
short-range, attractive interaction between cell-nuclei invoked in the
linear regime, we find a new effective force that emerges during the
non-linear regime. See Fig. 3.  This emergent force is
attractive close to the center and repulsive around the
edge. Since it is attractive near the center, the density of cell nuclei
is higher then near the edge. Where the emergent net force on cell
nuclei goes from attractive to repulsive is where we anticipate the
boundary of the cortex-core to be. If one were to invoke a Voronoi
tessellation of the cell nuclei to obtain cell shapes~\cite{Honda1978,Kaliman2016}, then the cell
shapes across this boundary would be elongated radially.
Given our continuum analysis, we cannot determine whether or not the
cortex is one cell layer thick or many cell layers thick.  Should
the cortex be many cell layers thick, then the cells farther away from
the boundary will not necessarily be elongated. In any event, it turns out
that the cortex is one cell layer thick, approximately.  The mechanism
for this phenomenon must be explored with a more detailed,
cellular-based model.  With a Voronoi tessellation, the positions of the
cell nuclei are the positions of the center of mass of a deformable
cell nuclei. If cells are elongated just beyond the zero-net force
boundary, then cell nuclei are as well since cell nuclei shape reflect
cell shape~\cite{Versaevel2012}. We, thus, infer the formation of a
cortex-core structure, though, again, the overall thickness of the
cortex has yet to be determined. We use the
term ``large-scale'' to denote that it is a multi-cellular
structure. 
In the case of
multiple cortex-core structures, supracellular actomyosin cabling~\cite{Yevick2019} may act as an
``external'' environment such that multiple cortex-core structure can
form simultaneously. Finally, for both the linear and non-linear
regimes, the cell nuclei densities were not large enough to worry
about overlaps between cell nuclei, or even shorter-range repulsive
interactions. 

\begin{figure*}
\includegraphics[width=0.8\textwidth]{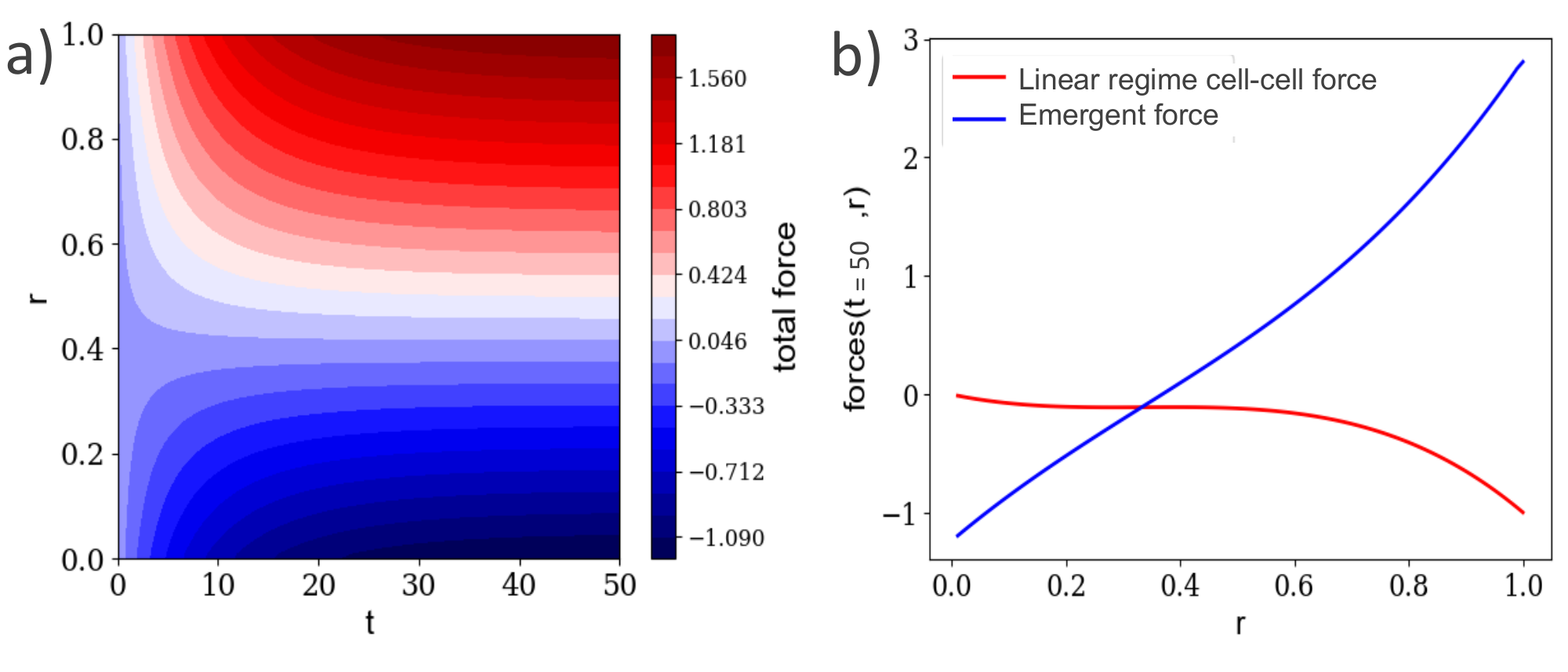}
\caption{a) The time evolution of the net force on the cell nuclei
  as a function of the radius of one cortex-core structure. 
 b) The emergent net force on cell nuclei at the final time and the initial net, attractive force between two cell nuclei used
 in the linear regime,   both as functions of the radius of one cortex-core structure.}
\lb{Fig:Forces}
\end{figure*}



\section{Foliation formation via a nonlinear buckling without bending model}
Now that cortex-core structures form, we proceed to the subsequent foliation
of the cortex observed in the ``2D A'' pathway. To do so, we turn to the BWBM model, which is a coarse-grained, continuum
model at a larger scale to accommodate the predominance of cell growth
at this stage. The
initial version of the BWBM model assumes a cortex-core structure and
has demonstrated qualitative agreement with
the foliation found in the quasi-two-dimensional brain organoids in pathway
``2D A'' with the
addition of blebbistatin~\cite{Engstrom2018}. A new 
nonlinearity, as we will show, extends the applicability of the model to the untreated
case. 

More precisely, we model the growing cortex-core structure as a two-dimensional
annulus-like region having outer radius $r$ and thickness $t$, which
are scalar functions of an angular coordinate $\theta$ such that $t$ is measured in the radial direction
(see Fig. 4). We also assume that $r$ and $t$ are single-valued, i.e., no
overhangs. We then introduce the quasi-static, coarse-grained energy functional
\begin{equation}
E[r,t,\tfrac{dt}{d\theta}] = \int d\theta \Big{\{} k_r(r-r_0)^2 -
k_t(t-t_0)^2 +\beta(1+\lambda t)\Big(\frac{dt}{d\theta}\Big)^2 \Big{\}},\label{Eq:BWBMM}
\end{equation}
to be minimized subject to a constraint on the area of the core, i.e.,
$\frac{1}{2}\int d\theta(r-t)^2=A_0=\textrm{constant}$. The variational problem then becomes 
$\delta\Big(E-\mu\int d\theta(r-t)^2\Big)=0,$
where $\mu$ is a Lagrange multiplier.  
\begin{figure*}
\centering
\includegraphics[width=0.8\textwidth]{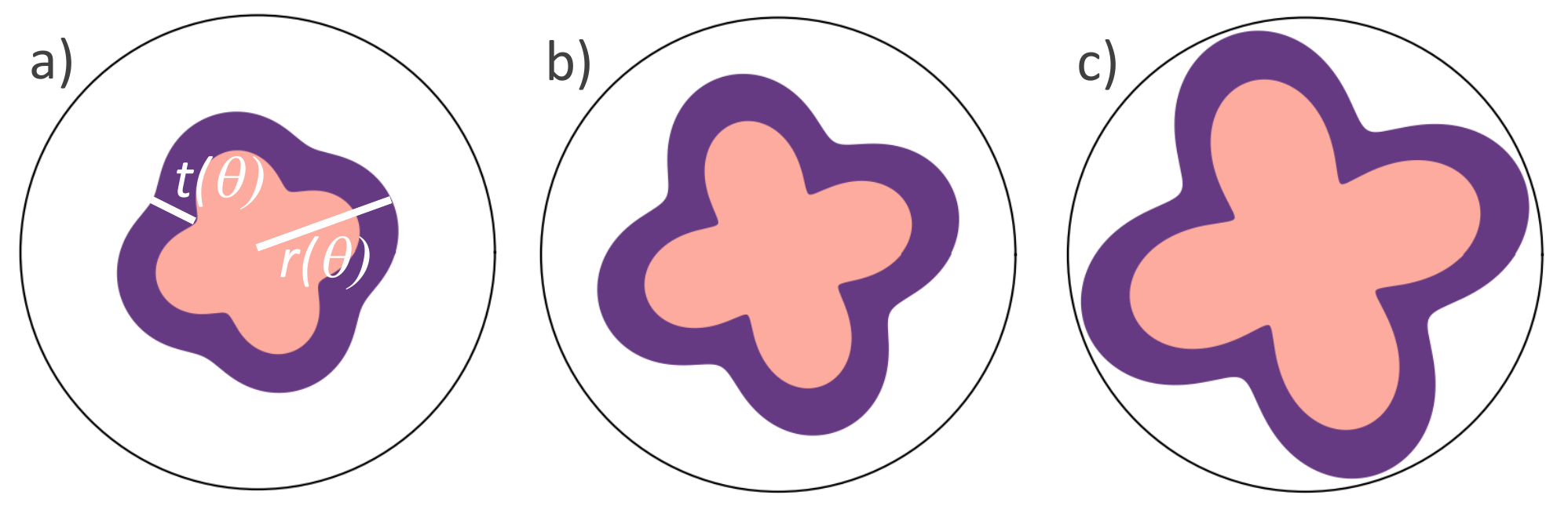}
\caption{BWBM model for the foliation of wild-type brain organoids
  pathway 2D A as time increases to the right. Parameters used are
  $k_t/\beta=15.6$, $t_0/r_0=0.7$, $\lambda/r_0=25$ for a)-c) and 
  for a) $\mu/k_r=0.72$, $k_r/k_t=0.043$, for b) $\mu/k_r=0.855$,
  $k_r/k_t=0.036$, for c) $\mu/k_r=0.9$, $k_r/k_t=0.034$.}
\end{figure*}

Shape change as a function of time, here, is encoded in
changes in the constants at hand. Addressing Eq.~\ref{Eq:BWBMM},
$k_r$, $k_t$, and $\beta$ are all
positive constants. The first term encodes a preferred radius $r_0$,
which we assume to be constant. This preferred
shape represents the energy cost in deforming the Matrigel, or the
extracellular environment. The second term favors thickening 
of the cortex with respect to a reference thickness $t_0$, which we
also assume to be constant, given its
negative contribution. Thus, while $k_r$ is a
modulus, $k_t$ can be regarded as a ``growth potential" in the form of
an anti-harmonic term. Therefore, the
validity domain of this analysis is only limited to those cases in
which the thicknesses are small, i.e., the energy functional is
bounded.

\begin{figure*}
\includegraphics[width=0.8\textwidth]{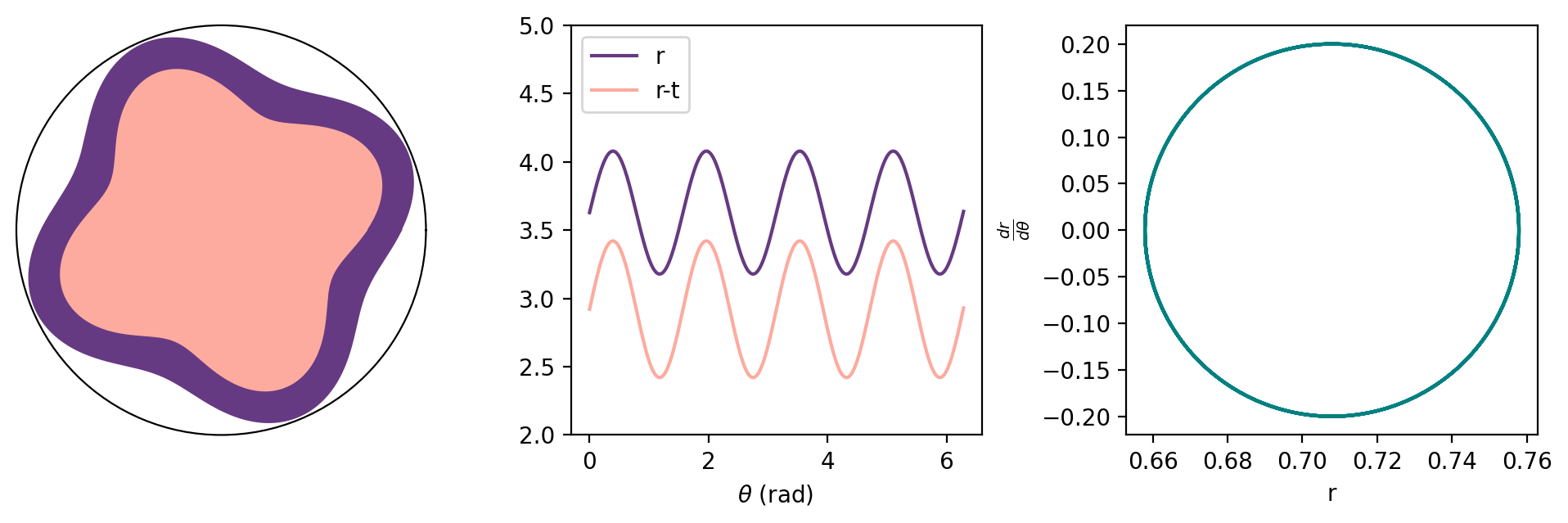}
\includegraphics[width=0.8\textwidth]{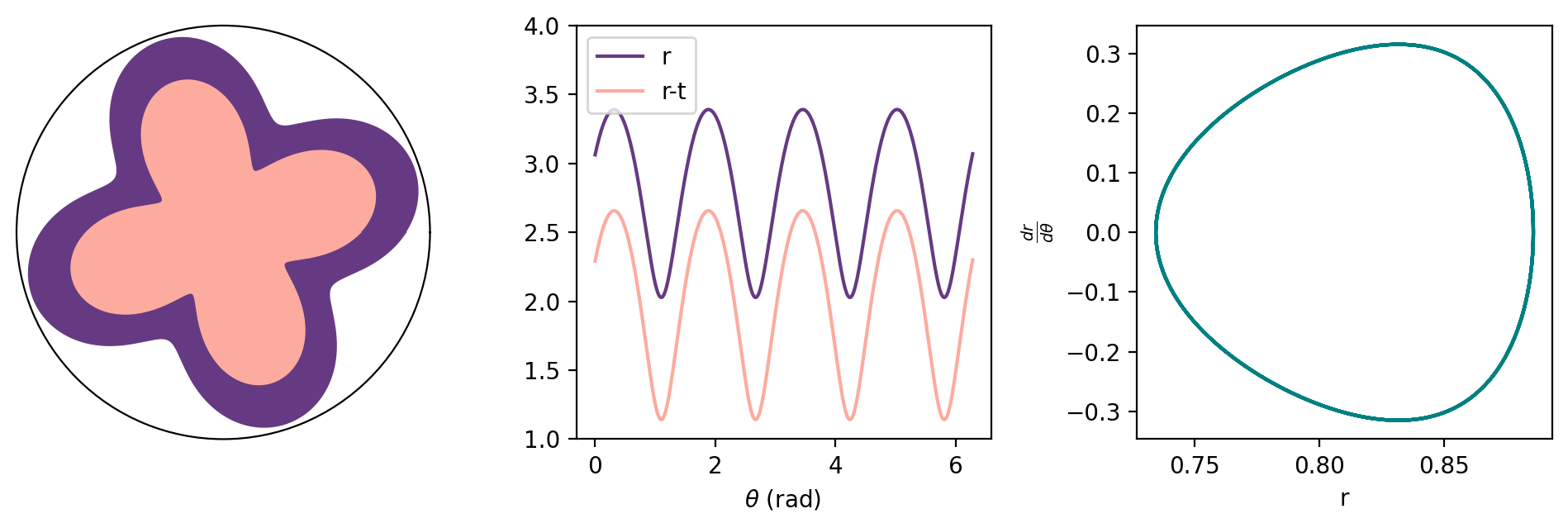}
\caption{Top row: Results for the linear version of the BWBM model, or
  $\lambda/r_0=0$, with other parameters the same as in Fig. 4c.
 Bottom row: Results for the nonlinear version of the BWBM model with 
  $\lambda/r_0=25$ for the first term in the perturbation expansion
  for the t-dependent mass term and with other parameters the same as in
Fig. 4c.}
\lb{Fig:BWBMcomparison}
\end{figure*}

The
corresponding terms compete with one another due to the 
incompressibility of the core, thereby driving the system away
from its preferred shape. The third term
penalizes spatial variations in thickness with the nonlinear $\lambda$
contribution representing  
the active, adaptive contractile nature of the cells.  As cortex cells are
extended/elongated, they build cytoskeletal
structures to adapt to the extension with the development of stress
fibers, for example, to regulate their strain and, therefore, resist the extension~\cite{Greiner2013}. The addition of blebbistatin prevents such structures
from resisting the extension, therefore, denoting the $\lambda=0$ case.  We note that
another form of nonlinearity has been studied in the context of
nonlinear elasticity of the radial glial cells spanning the
cerebellum~\cite{Gandikota2021}.  Here, there are no such radial glial
cells, at least during these early stages.

Assuming the initial cortex-core shape to be a circle with radius
$r_0$, the Euler-Lagrange equations result in
an unconventionally driven, nonlinear oscillator
equation. Specifically, the Euler-Lagrange equation for $t(\theta)$ is of
the form
\begin{equation}
(1+\lambda t)\frac{d^2t}{d\theta^2}+q^2t=-\frac{1}{2}\lambda (\frac{dt}{d\theta})^2+B,
\end{equation}
with $q^2=\frac{k_t}{\beta}[1+\frac{\epsilon c}{(1-\epsilon)}]$ and
$B=\frac{k_t}{\beta}[t_0+\frac{\epsilon c r_0}{ (1-\epsilon)}]$ after
defining $\epsilon=\frac{\mu}{k_r}$ and $c=\frac{k_r}{k_t}$. 
In addition, there is a linear relationship between $t$ and $r$, i.e.,
$r=\frac{-\epsilon t +r_0}{1-\epsilon}$. We
can, therefore, numerically solve for the shape of the cortex-core structure as a
function of the parameters. The RK45 method of the
\texttt{scipy.integrate} package in Python is used for the numerical
integration of the above nonlinear differential equation. 
Note that we treat the $t$-dependent mass term perturbatively given the existence of the usual, mass term.

The results for the subsequent brain organoid evolution are plotted in Fig. 4 for different $k_r$s, which decreases with time
as the Matrigel softens due to compression~\cite{VanOosten2016}.  We observe for the
nonlinear case an asymmetry developing between the crest (the gyri)
and the trough (the sulci) to
approach a more scalloped form prominent in the untreated brain
organoids.  To more clearly demonstrate the differences between the
linear BWBM model and this nonlinear version, we present shapes for
both cases in Fig.~\ref{Fig:BWBMcomparison}. On the other hand, the
scallops are not as packed tightly together as observed in the
experiments.  Interestingly, a recent nonlinear extension of the BWBM
model also demonstrated more scalloped foliation with a nonlinearity
introduced in $k_r$ to account for the nonlinear elasticity of the
radial glial cells~\cite{Gandikota2021}. Another interesting feature of the BWBM model is that once the first generation of foliations/folds appear, we anticipate the
potential for subsequent generations to occur as the boundaries of the
first generation foliation create a sub-system within the overall
structure such that foliation process can occur within the
sub-system, given the number of foliations is essentially
scale-invariant~\cite{Larsell,Engstrom2018,Gandikota2021}. In fact, this type of
higher-order branching process is observed in brain organoid experiments~\cite{Karzbrun2018} and in the developing, approximately
cylindrical cerebellum~\cite{Sudarov2007}.

\section{Discussion} 
We have established a two-part framework to quantify the shape of
brain organoids as they develop.  Both parts are rooted in the
assumption that the material is not purely elastic.
Indeed, tissue fluidity has emerged as a driver of shape change in
animal development more generally~\cite{mongera_2018,jain_2020}. The first part of the
framework models the interactions between cell nuclei due to activity
to examine how multiple, large-scale core-cortex
structures emerge in the confined case.  If we know the
initial density map of cell nuclei, we can predict the number and size of the cortex-core
structures.  We can
also predict the subsequent foliation of an individual cortex-core 
structure. Predictions for foliation in multiple cortex-core
structures require a more detailed analysis of an interacting version
of the BWBM model.

While Stage II of the ``2D B'' pathway was reported in the literature~\cite{Karzbrun2018}, it is not clear if such multi-core-cortex structures exhibit
Stage II behavior.  Considering just two-core-cortex structures with a very small
interface in between initially, then each cortex-core structure evolves
independently of each other until the interface increases due to the growth.  Earlier work has shown that the linear
BWBM model in the presence of a confining wall flattens the scallops~\cite{Gandikota2021}. Treating each structure as a confining presence of the
other, will thus, flatten the scallops and so one may observe some
foliation with different shapes along the interface between the two
structures as compared to the interface with Matrigel. However,
should the interface between the two structures not be small to begin
with, then one must also
treat the two core-cortex structures as a coupled system with the
spherical symmetry now broken.  We are currently extending the BWBM
model to describe multiple-cortex-core structures with interfaces in between, suggesting
that the extent of the foliation will depend on such details as the
difference in growth potentials between the two structures, etc.

While we have focused here on the structure of quasi-two-dimensional
brain organoids, three-dimensional brain organoid shapes typically consist of multiple
large-scale structures~\cite{Lancaster2013}. These large-scale, or multi-cellular structures are cortex-lumen
structures embedded within non-cortical/non-extended cells.  Our
framework for Stage I applies under these conditions as well.  Variations in
cell nuclear density,  as well as variations in
contractility of the cells,  determine where the
large-scale structures emerge.  Regions,  where the cellular contractility is less than
the average,  translate to cellular material effectively acting as a
passive, or extracellular, environment. The
regions of underdensity, as before, translate to an effective
repulsive force to push cells apart. The more the cells move apart,
the more likely ruptures will occur at the cell-cell interface to create a
lumen or hole.  The shape of such holes depends on the shape of the
regions of less active cells to which the more active cells are pulled
towards. Predicting the detailed shape of these large-scale
structures,  therefore,  requires some modeling at the cell-cell
interface level to pinpoint the rupture locations, which we do not
address here.  Recent work interpolating between
confluent and non-confluent tissue, thereby identifying points of
rupture, may help ~\cite{Kim2021}.  In the confined case, perhaps stronger interaction with the passive environment prevents such rupture.  However, our work suggests that a very homogenous
organoid with an underdensity region in the center of a spherical
organoid leads to one cortex-lumen structure embedded in a sea of
cells. See Fig.~\ref{Fig:StructuresWithHolesInThem}.

\begin{figure*}
\centering
\includegraphics[width=0.3\textwidth]{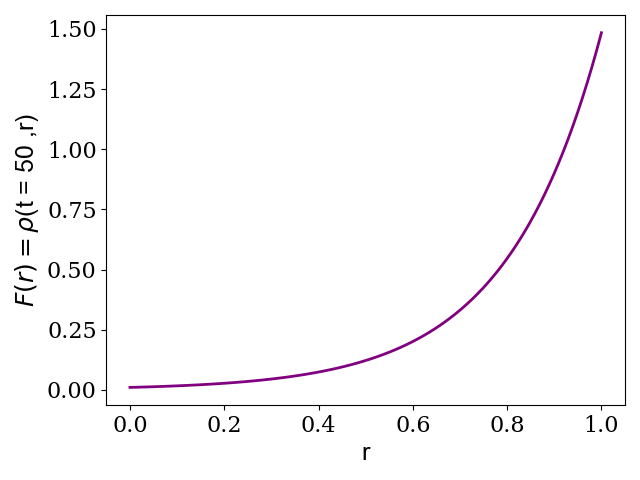}
\includegraphics[width=0.3\textwidth]{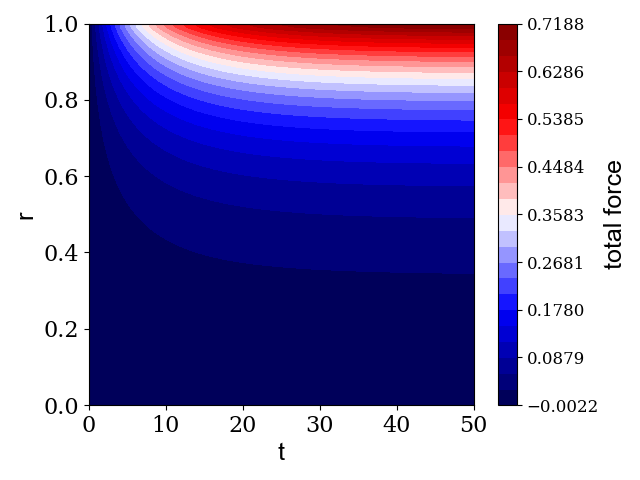}
\includegraphics[width=0.3\textwidth]{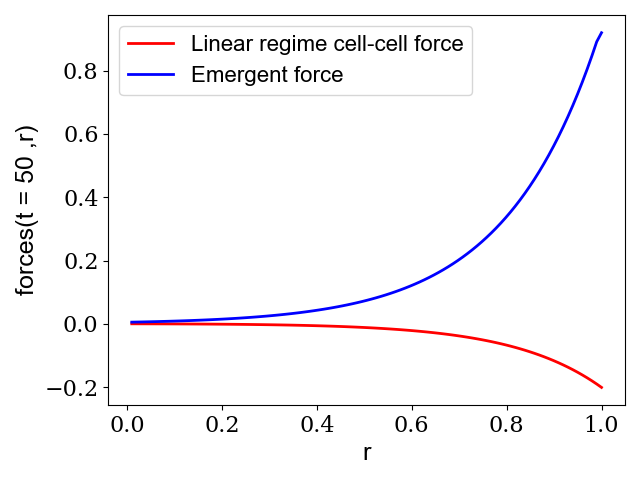}
\caption{In three dimensions, brain organoids have a hole, or lumen, at the center
  of large-scale/multi-cellular structure and the cell nuclei are all
  in the outer cortex region.  (Left): The number density
  of cell nuclei at the end of the nonlinear era, assuming a hole in the
  center of the cortex-lumen structure  (Middle): The time evolution
  of the net force on cell nuclei for a cortex-lumen structure.  
(Right): The initial, net attractive force on cell nuclei (red) due to 
cell-cell interactions and the emergent, net force (blue) as the cytoskeleton
restructures itself, using the assumed cell nuclei number density. 
\lb{Fig:StructuresWithHolesInThem}}
\end{figure*}

If we are to understand how the
brain attains its shape, brain organoids serve as an excellent {\it in
  vitro} platform.  While brain organoids do {\it not} currently mimic brain
shape, one can now design conditions in which higher-order foliations/folds are more
likely to occur in the confined case to more closely resemble the
cerebellum. Moreover, a three-dimensional brain organoid with
one cortex-lumen structure can potentially be engineered in which the cortex is
layered by the addition of cells at the cell-cell rupture site in
such a way that they also become extended.  Therefore, one can create more similar shapes to the mammalian brain or
even less similar to study how brain shape affects brain
function. For instance, the honeybee brain has a rather different structure
than a mammalian brain~\cite{Brandt2005}.  Moreover, at the heart of
subsequent brain development is the
existence of elongated cells that serve as a backbone for the
initiation of neurons, cells unique to the central nervous system,
which, therefore, requires more modeling attention. 

Finally, the framework we use to quantify large-scale
structure formation in brain organoids for Stage I is the same as the
hydrodynamic framework
for large-scale structure formation in the universe. In the early
universe, quantum fluctuations induce negligible mass
overdensities that grow over time, by attracting nearby mass, to
ultimately form galaxies
\cite{1982PhRvL..49.1110G,1982PhLB..117..175S,1982PhLB..115..295H}. Otherwise,
despite the attractive nature of gravity, an exactly uniform universe will stay
uniform forever. Given
this mathematical correspondence, perhaps studying multi-cellular 
structure formation in a petri dish with living matter may tell us
something intriguing about the potential for engineering new types of
mini-universes, i.e., large-scale structure formation in these new kinds of universes and morphogenesis are inextricably linked. 

\begin{acknowledgments}
JMS acknowledges Mahesh Gandikota and Orly Reiner for discussions and
finanical support from grant NSF-DMR-1832002
and an Isaac Newton Award from the DoD. 
\end{acknowledgments}

\bibliography{brainshape}

\begin{thebibliography}{43}%
\makeatletter
\providecommand \@ifxundefined [1]{%
 \@ifx{#1\undefined}
}%
\providecommand \@ifnum [1]{%
 \ifnum #1\expandafter \@firstoftwo
 \else \expandafter \@secondoftwo
 \fi
}%
\providecommand \@ifx [1]{%
 \ifx #1\expandafter \@firstoftwo
 \else \expandafter \@secondoftwo
 \fi
}%
\providecommand \natexlab [1]{#1}%
\providecommand \enquote  [1]{``#1''}%
\providecommand \bibnamefont  [1]{#1}%
\providecommand \bibfnamefont [1]{#1}%
\providecommand \citenamefont [1]{#1}%
\providecommand \href@noop [0]{\@secondoftwo}%
\providecommand \href [0]{\begingroup \@sanitize@url \@href}%
\providecommand \@href[1]{\@@startlink{#1}\@@href}%
\providecommand \@@href[1]{\endgroup#1\@@endlink}%
\providecommand \@sanitize@url [0]{\catcode `\\12\catcode `\$12\catcode
  `\&12\catcode `\#12\catcode `\^12\catcode `\_12\catcode `\%12\relax}%
\providecommand \@@startlink[1]{}%
\providecommand \@@endlink[0]{}%
\providecommand \url  [0]{\begingroup\@sanitize@url \@url }%
\providecommand \@url [1]{\endgroup\@href {#1}{\urlprefix }}%
\providecommand \urlprefix  [0]{URL }%
\providecommand \Eprint [0]{\href }%
\providecommand \doibase [0]{https://doi.org/}%
\providecommand \selectlanguage [0]{\@gobble}%
\providecommand \bibinfo  [0]{\@secondoftwo}%
\providecommand \bibfield  [0]{\@secondoftwo}%
\providecommand \translation [1]{[#1]}%
\providecommand \BibitemOpen [0]{}%
\providecommand \bibitemStop [0]{}%
\providecommand \bibitemNoStop [0]{.\EOS\space}%
\providecommand \EOS [0]{\spacefactor3000\relax}%
\providecommand \BibitemShut  [1]{\csname bibitem#1\endcsname}%
\let\auto@bib@innerbib\@empty
\bibitem [{\citenamefont {Sanes}\ \emph {et~al.}(2012)\citenamefont {Sanes},
  \citenamefont {Reh},\ and\ \citenamefont {Harris}}]{DevNervousSystemBook}%
  \BibitemOpen
  \bibfield  {author} {\bibinfo {author} {\bibfnamefont {D.~H.}\ \bibnamefont
  {Sanes}}, \bibinfo {author} {\bibfnamefont {T.~A.}\ \bibnamefont {Reh}},\
  and\ \bibinfo {author} {\bibfnamefont {W.~A.}\ \bibnamefont {Harris}},\
  }\href@noop {} {\emph {\bibinfo {title} {Development of the Nervous
  System}}}\ (\bibinfo  {publisher} {Academic Press},\ \bibinfo {year}
  {2012})\BibitemShut {NoStop}%
\bibitem [{\citenamefont {Richman}\ \emph {et~al.}(1975)\citenamefont
  {Richman}, \citenamefont {Stewart}, \citenamefont {Hutchinson},\ and\
  \citenamefont {Caviness}}]{Richman1975}%
  \BibitemOpen
  \bibfield  {author} {\bibinfo {author} {\bibfnamefont {D.~P.}\ \bibnamefont
  {Richman}}, \bibinfo {author} {\bibfnamefont {R.~M.}\ \bibnamefont
  {Stewart}}, \bibinfo {author} {\bibfnamefont {J.~W.}\ \bibnamefont
  {Hutchinson}},\ and\ \bibinfo {author} {\bibfnamefont {V.~S.}\ \bibnamefont
  {Caviness}},\ }\bibfield  {title} {\bibinfo {title} {Mechanical model of
  brain convolutional development},\ }\href@noop {} {\bibfield  {journal}
  {\bibinfo  {journal} {Science}\ }\textbf {\bibinfo {volume} {189}},\ \bibinfo
  {pages} {18} (\bibinfo {year} {1975})}\BibitemShut {NoStop}%
\bibitem [{\citenamefont {Van~Essen}(1997)}]{VanEssen1997}%
  \BibitemOpen
  \bibfield  {author} {\bibinfo {author} {\bibfnamefont {D.~C.}\ \bibnamefont
  {Van~Essen}},\ }\bibfield  {title} {\bibinfo {title} {A tension-based theory
  of morphogenesis and compact wiring in the central nervous system},\
  }\href@noop {} {\bibfield  {journal} {\bibinfo  {journal} {Nature}\ }\textbf
  {\bibinfo {volume} {385}},\ \bibinfo {pages} {313} (\bibinfo {year}
  {1997})}\BibitemShut {NoStop}%
\bibitem [{\citenamefont {Bayly}\ \emph {et~al.}(2013)\citenamefont {Bayly},
  \citenamefont {Okamoto}, \citenamefont {Xu}, \citenamefont {Shi},\ and\
  \citenamefont {Taber}}]{Bayly2013}%
  \BibitemOpen
  \bibfield  {author} {\bibinfo {author} {\bibfnamefont {P.}~\bibnamefont
  {Bayly}}, \bibinfo {author} {\bibfnamefont {R.}~\bibnamefont {Okamoto}},
  \bibinfo {author} {\bibfnamefont {G.}~\bibnamefont {Xu}}, \bibinfo {author}
  {\bibfnamefont {Y.}~\bibnamefont {Shi}},\ and\ \bibinfo {author}
  {\bibfnamefont {L.}~\bibnamefont {Taber}},\ }\bibfield  {title} {\bibinfo
  {title} {A cortical folding model incorporating stress-dependent growth
  explains gyral wavelengths and stress patterns in the developing brain},\
  }\href@noop {} {\bibfield  {journal} {\bibinfo  {journal} {Physical Biology}\
  }\textbf {\bibinfo {volume} {10}},\ \bibinfo {pages} {016005} (\bibinfo
  {year} {2013})}\BibitemShut {NoStop}%
\bibitem [{\citenamefont {Manyuhina}\ \emph {et~al.}(2014)\citenamefont
  {Manyuhina}, \citenamefont {Mayett},\ and\ \citenamefont
  {Schwarz}}]{Manyuhina2014}%
  \BibitemOpen
  \bibfield  {author} {\bibinfo {author} {\bibfnamefont {O.~V.}\ \bibnamefont
  {Manyuhina}}, \bibinfo {author} {\bibfnamefont {D.}~\bibnamefont {Mayett}},\
  and\ \bibinfo {author} {\bibfnamefont {J.~M.}\ \bibnamefont {Schwarz}},\
  }\bibfield  {title} {\bibinfo {title} {Elastic instabilities in a layered
  cerebral cortex: a revised axonal tension model for cortex folding},\
  }\href@noop {} {\bibfield  {journal} {\bibinfo  {journal} {New Journal of
  Physics}\ }\textbf {\bibinfo {volume} {16}},\ \bibinfo {pages} {123058}
  (\bibinfo {year} {2014})}\BibitemShut {NoStop}%
\bibitem [{\citenamefont {Budday}\ \emph {et~al.}(2015)\citenamefont {Budday},
  \citenamefont {Steinmann~III},\ and\ \citenamefont {Kuhl}}]{Budday2015}%
  \BibitemOpen
  \bibfield  {author} {\bibinfo {author} {\bibfnamefont {S.}~\bibnamefont
  {Budday}}, \bibinfo {author} {\bibfnamefont {P.}~\bibnamefont
  {Steinmann~III}},\ and\ \bibinfo {author} {\bibfnamefont {E.}~\bibnamefont
  {Kuhl}},\ }\bibfield  {title} {\bibinfo {title} {Physical biology of human
  brain development},\ }\href@noop {} {\bibfield  {journal} {\bibinfo
  {journal} {Frontiers in Cellular Neuroscience}\ }\textbf {\bibinfo {volume}
  {9}},\ \bibinfo {pages} {257} (\bibinfo {year} {2015})}\BibitemShut {NoStop}%
\bibitem [{\citenamefont {Mota}\ and\ \citenamefont
  {Herculano-Houzel}(2015)}]{Mota2015}%
  \BibitemOpen
  \bibfield  {author} {\bibinfo {author} {\bibfnamefont {B.}~\bibnamefont
  {Mota}}\ and\ \bibinfo {author} {\bibfnamefont {S.}~\bibnamefont
  {Herculano-Houzel}},\ }\bibfield  {title} {\bibinfo {title} {Cortical folding
  scales universally with surface area and thickness, not number of neurons},\
  }\href {https://doi.org/10.1126/science.aaa9101} {\bibfield  {journal}
  {\bibinfo  {journal} {Science}\ }\textbf {\bibinfo {volume} {349}},\ \bibinfo
  {pages} {74} (\bibinfo {year} {2015})}\BibitemShut {NoStop}%
\bibitem [{\citenamefont {Tallinen}\ \emph {et~al.}(2016)\citenamefont
  {Tallinen}, \citenamefont {Chung}, \citenamefont {Rousseau}, \citenamefont
  {Girard}, \citenamefont {Lef{\`e}vre},\ and\ \citenamefont
  {Mahadevan}}]{Tallinen2016}%
  \BibitemOpen
  \bibfield  {author} {\bibinfo {author} {\bibfnamefont {T.}~\bibnamefont
  {Tallinen}}, \bibinfo {author} {\bibfnamefont {J.~Y.}\ \bibnamefont {Chung}},
  \bibinfo {author} {\bibfnamefont {F.}~\bibnamefont {Rousseau}}, \bibinfo
  {author} {\bibfnamefont {N.}~\bibnamefont {Girard}}, \bibinfo {author}
  {\bibfnamefont {J.}~\bibnamefont {Lef{\`e}vre}},\ and\ \bibinfo {author}
  {\bibfnamefont {L.}~\bibnamefont {Mahadevan}},\ }\bibfield  {title} {\bibinfo
  {title} {On the growth and form of cortical convolutions},\ }\href@noop {}
  {\bibfield  {journal} {\bibinfo  {journal} {Nature Physics}\ }\textbf
  {\bibinfo {volume} {12}},\ \bibinfo {pages} {588} (\bibinfo {year}
  {2016})}\BibitemShut {NoStop}%
\bibitem [{\citenamefont {Lejeune}\ \emph {et~al.}(2016)\citenamefont
  {Lejeune}, \citenamefont {Javili}, \citenamefont {Weickenmeier},
  \citenamefont {Kuhl},\ and\ \citenamefont {Linder}}]{Lejeune2016}%
  \BibitemOpen
  \bibfield  {author} {\bibinfo {author} {\bibfnamefont {E.}~\bibnamefont
  {Lejeune}}, \bibinfo {author} {\bibfnamefont {A.}~\bibnamefont {Javili}},
  \bibinfo {author} {\bibfnamefont {J.}~\bibnamefont {Weickenmeier}}, \bibinfo
  {author} {\bibfnamefont {E.}~\bibnamefont {Kuhl}},\ and\ \bibinfo {author}
  {\bibfnamefont {C.}~\bibnamefont {Linder}},\ }\bibfield  {title} {\bibinfo
  {title} {Tri-layer wrinkling as a mechanism for anchoring center initiation
  in the developing cerebellum},\ }\href@noop {} {\bibfield  {journal}
  {\bibinfo  {journal} {Soft Matter}\ }\textbf {\bibinfo {volume} {12}},\
  \bibinfo {pages} {5613} (\bibinfo {year} {2016})}\BibitemShut {NoStop}%
\bibitem [{\citenamefont {Lejeune}\ \emph {et~al.}(2019)\citenamefont
  {Lejeune}, \citenamefont {Dortdivanlioglu}, \citenamefont {Kuhl},\ and\
  \citenamefont {Linder}}]{Lejeune2019}%
  \BibitemOpen
  \bibfield  {author} {\bibinfo {author} {\bibfnamefont {E.}~\bibnamefont
  {Lejeune}}, \bibinfo {author} {\bibfnamefont {B.}~\bibnamefont
  {Dortdivanlioglu}}, \bibinfo {author} {\bibfnamefont {E.}~\bibnamefont
  {Kuhl}},\ and\ \bibinfo {author} {\bibfnamefont {C.}~\bibnamefont {Linder}},\
  }\bibfield  {title} {\bibinfo {title} {Understanding the mechanical link
  between oriented cell division and cerebellar morphogenesis},\ }\href@noop {}
  {\bibfield  {journal} {\bibinfo  {journal} {Soft Matter}\ }\textbf {\bibinfo
  {volume} {15}},\ \bibinfo {pages} {2204} (\bibinfo {year}
  {2019})}\BibitemShut {NoStop}%
\bibitem [{\citenamefont {Engstrom}\ \emph {et~al.}(2018)\citenamefont
  {Engstrom}, \citenamefont {Zhang}, \citenamefont {Lawton}, \citenamefont
  {Joyner},\ and\ \citenamefont {Schwarz}}]{Engstrom2018}%
  \BibitemOpen
  \bibfield  {author} {\bibinfo {author} {\bibfnamefont {T.}~\bibnamefont
  {Engstrom}}, \bibinfo {author} {\bibfnamefont {T.}~\bibnamefont {Zhang}},
  \bibinfo {author} {\bibfnamefont {A.}~\bibnamefont {Lawton}}, \bibinfo
  {author} {\bibfnamefont {A.}~\bibnamefont {Joyner}},\ and\ \bibinfo {author}
  {\bibfnamefont {J.~M.}\ \bibnamefont {Schwarz}},\ }\bibfield  {title}
  {\bibinfo {title} {Buckling without bending: a new paradigm in
  morphogenesis},\ }\href@noop {} {\bibfield  {journal} {\bibinfo  {journal}
  {Physical Review X}\ }\textbf {\bibinfo {volume} {8}},\ \bibinfo {pages}
  {041053} (\bibinfo {year} {2018})}\BibitemShut {NoStop}%
\bibitem [{\citenamefont {Lawton}\ \emph {et~al.}(2019)\citenamefont {Lawton},
  \citenamefont {Engstrom}, \citenamefont {Rohrbach}, \citenamefont {Omura},
  \citenamefont {Turnbull}, \citenamefont {Mamou}, \citenamefont {Zhang},
  \citenamefont {Schwarz},\ and\ \citenamefont {Joyner}}]{Lawton2019}%
  \BibitemOpen
  \bibfield  {author} {\bibinfo {author} {\bibfnamefont {A.~K.}\ \bibnamefont
  {Lawton}}, \bibinfo {author} {\bibfnamefont {T.}~\bibnamefont {Engstrom}},
  \bibinfo {author} {\bibfnamefont {D.}~\bibnamefont {Rohrbach}}, \bibinfo
  {author} {\bibfnamefont {M.}~\bibnamefont {Omura}}, \bibinfo {author}
  {\bibfnamefont {D.~H.}\ \bibnamefont {Turnbull}}, \bibinfo {author}
  {\bibfnamefont {J.}~\bibnamefont {Mamou}}, \bibinfo {author} {\bibfnamefont
  {T.}~\bibnamefont {Zhang}}, \bibinfo {author} {\bibfnamefont {J.~M.}\
  \bibnamefont {Schwarz}},\ and\ \bibinfo {author} {\bibfnamefont {A.~L.}\
  \bibnamefont {Joyner}},\ }\bibfield  {title} {\bibinfo {title} {Cerebellar
  folding is initiated by mechanical constraints on a fluid-like layer without
  a cellular pre-pattern},\ }\href@noop {} {\bibfield  {journal} {\bibinfo
  {journal} {Elife}\ }\textbf {\bibinfo {volume} {8}},\ \bibinfo {pages}
  {e45019} (\bibinfo {year} {2019})}\BibitemShut {NoStop}%
\bibitem [{\citenamefont {Xu}\ \emph {et~al.}(2010)\citenamefont {Xu},
  \citenamefont {Knutsen}, \citenamefont {Dikranian}, \citenamefont {Bayly},\
  and\ \citenamefont {A.}}]{Xu2010}%
  \BibitemOpen
  \bibfield  {author} {\bibinfo {author} {\bibfnamefont {G.}~\bibnamefont
  {Xu}}, \bibinfo {author} {\bibfnamefont {A.~K.}\ \bibnamefont {Knutsen}},
  \bibinfo {author} {\bibfnamefont {C.~D.}\ \bibnamefont {Dikranian},
  \bibfnamefont {K.~Kroenke}}, \bibinfo {author} {\bibfnamefont {P.~V.}\
  \bibnamefont {Bayly}},\ and\ \bibinfo {author} {\bibfnamefont {T.~L.}\
  \bibnamefont {A.}},\ }\bibfield  {title} {\bibinfo {title} {Axons pull on the
  brain, but tension does not drive cortical folding.},\ }\href@noop {}
  {\bibfield  {journal} {\bibinfo  {journal} {J. Biomech. Eng.}\ }\textbf
  {\bibinfo {volume} {132}},\ \bibinfo {pages} {071013} (\bibinfo {year}
  {2010})}\BibitemShut {NoStop}%
\bibitem [{\citenamefont {Lancaster}\ \emph {et~al.}(2013)\citenamefont
  {Lancaster}, \citenamefont {Renner}, \citenamefont {Martin}, \citenamefont
  {Wenzel}, \citenamefont {Bicknell}, \citenamefont {Hurles}, \citenamefont
  {Homfray}, \citenamefont {Penninger}, \citenamefont {Jackson},\ and\
  \citenamefont {Knoblich}}]{Lancaster2013}%
  \BibitemOpen
  \bibfield  {author} {\bibinfo {author} {\bibfnamefont {M.~A.}\ \bibnamefont
  {Lancaster}}, \bibinfo {author} {\bibfnamefont {M.}~\bibnamefont {Renner}},
  \bibinfo {author} {\bibfnamefont {C.-A.}\ \bibnamefont {Martin}}, \bibinfo
  {author} {\bibfnamefont {D.}~\bibnamefont {Wenzel}}, \bibinfo {author}
  {\bibfnamefont {L.~S.}\ \bibnamefont {Bicknell}}, \bibinfo {author}
  {\bibfnamefont {M.~E.}\ \bibnamefont {Hurles}}, \bibinfo {author}
  {\bibfnamefont {T.}~\bibnamefont {Homfray}}, \bibinfo {author} {\bibfnamefont
  {J.~M.}\ \bibnamefont {Penninger}}, \bibinfo {author} {\bibfnamefont {A.~P.}\
  \bibnamefont {Jackson}},\ and\ \bibinfo {author} {\bibfnamefont {J.~A.}\
  \bibnamefont {Knoblich}},\ }\bibfield  {title} {\bibinfo {title} {Cerebral
  organoids model human brain development and microcephaly},\ }\href@noop {}
  {\bibfield  {journal} {\bibinfo  {journal} {Nature}\ }\textbf {\bibinfo
  {volume} {501}},\ \bibinfo {pages} {373} (\bibinfo {year}
  {2013})}\BibitemShut {NoStop}%
\bibitem [{\citenamefont {Benito-Kwiecinski}\ \emph {et~al.}(2021)\citenamefont
  {Benito-Kwiecinski}, \citenamefont {Giandomenico}, \citenamefont {Sutcliffe},
  \citenamefont {Riis}, \citenamefont {Freire-Pritchett}, \citenamefont
  {Kelava}, \citenamefont {Wunderlich}, \citenamefont {Martin}, \citenamefont
  {Wray}, \citenamefont {McDole},\ and\ \citenamefont
  {Lancaster}}]{Lancaster2021}%
  \BibitemOpen
  \bibfield  {author} {\bibinfo {author} {\bibfnamefont {S.}~\bibnamefont
  {Benito-Kwiecinski}}, \bibinfo {author} {\bibfnamefont {S.~L.}\ \bibnamefont
  {Giandomenico}}, \bibinfo {author} {\bibfnamefont {M.}~\bibnamefont
  {Sutcliffe}}, \bibinfo {author} {\bibfnamefont {E.~S.}\ \bibnamefont {Riis}},
  \bibinfo {author} {\bibfnamefont {P.}~\bibnamefont {Freire-Pritchett}},
  \bibinfo {author} {\bibfnamefont {I.}~\bibnamefont {Kelava}}, \bibinfo
  {author} {\bibfnamefont {S.}~\bibnamefont {Wunderlich}}, \bibinfo {author}
  {\bibfnamefont {U.}~\bibnamefont {Martin}}, \bibinfo {author} {\bibfnamefont
  {G.~A.}\ \bibnamefont {Wray}}, \bibinfo {author} {\bibfnamefont
  {K.}~\bibnamefont {McDole}},\ and\ \bibinfo {author} {\bibfnamefont {M.~A.}\
  \bibnamefont {Lancaster}},\ }\bibfield  {title} {\bibinfo {title} {An early
  cell shape transition drives evolutionary expansion of the human forebrain},\
  }\href@noop {} {\bibfield  {journal} {\bibinfo  {journal} {Cell}\ }\textbf
  {\bibinfo {volume} {184}},\ \bibinfo {pages} {2084} (\bibinfo {year}
  {2021})}\BibitemShut {NoStop}%
\bibitem [{\citenamefont {Karzbrun}\ \emph {et~al.}(2018)\citenamefont
  {Karzbrun}, \citenamefont {Kshirsagar}, \citenamefont {Cohen}, \citenamefont
  {Hanna},\ and\ \citenamefont {Reiner}}]{Karzbrun2018}%
  \BibitemOpen
  \bibfield  {author} {\bibinfo {author} {\bibfnamefont {E.}~\bibnamefont
  {Karzbrun}}, \bibinfo {author} {\bibfnamefont {A.}~\bibnamefont
  {Kshirsagar}}, \bibinfo {author} {\bibfnamefont {S.~R.}\ \bibnamefont
  {Cohen}}, \bibinfo {author} {\bibfnamefont {J.~H.}\ \bibnamefont {Hanna}},\
  and\ \bibinfo {author} {\bibfnamefont {O.}~\bibnamefont {Reiner}},\
  }\bibfield  {title} {\bibinfo {title} {Human brain organoids on a chip reveal
  the physics of folding},\ }\href@noop {} {\bibfield  {journal} {\bibinfo
  {journal} {Nature Physics}\ }\textbf {\bibinfo {volume} {14}},\ \bibinfo
  {pages} {515} (\bibinfo {year} {2018})}\BibitemShut {NoStop}%
\bibitem [{\citenamefont {Velasco}\ \emph {et~al.}(2019)\citenamefont
  {Velasco}, \citenamefont {Kedaigle}, \citenamefont {Simmons}, \citenamefont
  {Nash}, \citenamefont {Rocha}, \citenamefont {Quadrato}, \citenamefont
  {Paulsen}, \citenamefont {Nguyen}, \citenamefont {Adiconis}, \citenamefont
  {Regev}, \citenamefont {Levin},\ and\ \citenamefont {Arlotta}}]{Velasco2019}%
  \BibitemOpen
  \bibfield  {author} {\bibinfo {author} {\bibfnamefont {S.}~\bibnamefont
  {Velasco}}, \bibinfo {author} {\bibfnamefont {A.~J.}\ \bibnamefont
  {Kedaigle}}, \bibinfo {author} {\bibfnamefont {S.~K.}\ \bibnamefont
  {Simmons}}, \bibinfo {author} {\bibfnamefont {A.}~\bibnamefont {Nash}},
  \bibinfo {author} {\bibfnamefont {M.}~\bibnamefont {Rocha}}, \bibinfo
  {author} {\bibfnamefont {G.}~\bibnamefont {Quadrato}}, \bibinfo {author}
  {\bibfnamefont {B.}~\bibnamefont {Paulsen}}, \bibinfo {author} {\bibfnamefont
  {L.}~\bibnamefont {Nguyen}}, \bibinfo {author} {\bibfnamefont
  {X.}~\bibnamefont {Adiconis}}, \bibinfo {author} {\bibfnamefont
  {A.}~\bibnamefont {Regev}}, \bibinfo {author} {\bibfnamefont {J.~Z.}\
  \bibnamefont {Levin}},\ and\ \bibinfo {author} {\bibfnamefont
  {P.}~\bibnamefont {Arlotta}},\ }\bibfield  {title} {\bibinfo {title}
  {Individual brain organoids reproducibly form cell diversity of the human
  cerebral cortex},\ }\href@noop {} {\bibfield  {journal} {\bibinfo  {journal}
  {Nature}\ }\textbf {\bibinfo {volume} {570}},\ \bibinfo {pages} {523}
  (\bibinfo {year} {2019})}\BibitemShut {NoStop}%
\bibitem [{\citenamefont {Trujillo}\ \emph {et~al.}(2019)\citenamefont
  {Trujillo}, \citenamefont {Gao}, \citenamefont {Negraes},\ and\ \citenamefont
  {et~al.}}]{Trujillo2019}%
  \BibitemOpen
  \bibfield  {author} {\bibinfo {author} {\bibfnamefont {C.~A.}\ \bibnamefont
  {Trujillo}}, \bibinfo {author} {\bibfnamefont {R.}~\bibnamefont {Gao}},
  \bibinfo {author} {\bibfnamefont {P.~D.}\ \bibnamefont {Negraes}},\ and\
  \bibinfo {author} {\bibnamefont {et~al.}},\ }\bibfield  {title} {\bibinfo
  {title} {Complex oscillatory waves emerging from cortical organoids model
  early human brain network development},\ }\href@noop {} {\bibfield  {journal}
  {\bibinfo  {journal} {Cell Stem Cell}\ }\textbf {\bibinfo {volume} {25}},\
  \bibinfo {pages} {558} (\bibinfo {year} {2019})}\BibitemShut {NoStop}%
\bibitem [{\citenamefont {Beaune}\ \emph {et~al.}(2014)\citenamefont {Beaune},
  \citenamefont {Stirbat}, \citenamefont {Khalifat}, \citenamefont
  {Cochet-Escartin}, \citenamefont {Garcia}, \citenamefont {Gurchenkov},
  \citenamefont {Murrell}, \citenamefont {Dufour}, \citenamefont {Cuvelier},\
  and\ \citenamefont {Brochard-Wyart}}]{Beaune8055}%
  \BibitemOpen
  \bibfield  {author} {\bibinfo {author} {\bibfnamefont {G.}~\bibnamefont
  {Beaune}}, \bibinfo {author} {\bibfnamefont {T.~V.}\ \bibnamefont {Stirbat}},
  \bibinfo {author} {\bibfnamefont {N.}~\bibnamefont {Khalifat}}, \bibinfo
  {author} {\bibfnamefont {O.}~\bibnamefont {Cochet-Escartin}}, \bibinfo
  {author} {\bibfnamefont {S.}~\bibnamefont {Garcia}}, \bibinfo {author}
  {\bibfnamefont {V.~V.}\ \bibnamefont {Gurchenkov}}, \bibinfo {author}
  {\bibfnamefont {M.~P.}\ \bibnamefont {Murrell}}, \bibinfo {author}
  {\bibfnamefont {S.}~\bibnamefont {Dufour}}, \bibinfo {author} {\bibfnamefont
  {D.}~\bibnamefont {Cuvelier}},\ and\ \bibinfo {author} {\bibfnamefont
  {F.}~\bibnamefont {Brochard-Wyart}},\ }\bibfield  {title} {\bibinfo {title}
  {How cells flow in the spreading of cellular aggregates},\ }\href
  {https://doi.org/10.1073/pnas.1323788111} {\bibfield  {journal} {\bibinfo
  {journal} {Proceedings of the National Academy of Sciences}\ }\textbf
  {\bibinfo {volume} {111}},\ \bibinfo {pages} {8055} (\bibinfo {year}
  {2014})},\ \Eprint
  {https://arxiv.org/abs/https://www.pnas.org/content/111/22/8055.full.pdf}
  {https://www.pnas.org/content/111/22/8055.full.pdf} \BibitemShut {NoStop}%
\bibitem [{\citenamefont {Stirbat}\ \emph {et~al.}(2013)\citenamefont
  {Stirbat}, \citenamefont {Mgharbel}, \citenamefont {Bodennec}, \citenamefont
  {Ferri}, \citenamefont {Mertani}, \citenamefont {Rieu},\ and\ \citenamefont
  {Delano{\"e}-Ayari}}]{stirbat2013fine}%
  \BibitemOpen
  \bibfield  {author} {\bibinfo {author} {\bibfnamefont {T.~V.}\ \bibnamefont
  {Stirbat}}, \bibinfo {author} {\bibfnamefont {A.}~\bibnamefont {Mgharbel}},
  \bibinfo {author} {\bibfnamefont {S.}~\bibnamefont {Bodennec}}, \bibinfo
  {author} {\bibfnamefont {K.}~\bibnamefont {Ferri}}, \bibinfo {author}
  {\bibfnamefont {H.~C.}\ \bibnamefont {Mertani}}, \bibinfo {author}
  {\bibfnamefont {J.-P.}\ \bibnamefont {Rieu}},\ and\ \bibinfo {author}
  {\bibfnamefont {H.}~\bibnamefont {Delano{\"e}-Ayari}},\ }\bibfield  {title}
  {\bibinfo {title} {Fine tuning of tissues' viscosity and surface tension
  through contractility suggests a new role for $\alpha$-catenin},\ }\href@noop
  {} {\bibfield  {journal} {\bibinfo  {journal} {PloS one}\ }\textbf {\bibinfo
  {volume} {8}},\ \bibinfo {pages} {e52554} (\bibinfo {year}
  {2013})}\BibitemShut {NoStop}%
\bibitem [{\citenamefont {Guevorkian}\ \emph {et~al.}(2010)\citenamefont
  {Guevorkian}, \citenamefont {Colbert}, \citenamefont {Durth}, \citenamefont
  {Dufour},\ and\ \citenamefont {Brochard-Wyart}}]{PhysRevLett.104.218101}%
  \BibitemOpen
  \bibfield  {author} {\bibinfo {author} {\bibfnamefont {K.}~\bibnamefont
  {Guevorkian}}, \bibinfo {author} {\bibfnamefont {M.-J.}\ \bibnamefont
  {Colbert}}, \bibinfo {author} {\bibfnamefont {M.}~\bibnamefont {Durth}},
  \bibinfo {author} {\bibfnamefont {S.}~\bibnamefont {Dufour}},\ and\ \bibinfo
  {author} {\bibfnamefont {F.~m.~c.}\ \bibnamefont {Brochard-Wyart}},\
  }\bibfield  {title} {\bibinfo {title} {Aspiration of biological viscoelastic
  drops},\ }\href {https://doi.org/10.1103/PhysRevLett.104.218101} {\bibfield
  {journal} {\bibinfo  {journal} {Phys. Rev. Lett.}\ }\textbf {\bibinfo
  {volume} {104}},\ \bibinfo {pages} {218101} (\bibinfo {year}
  {2010})}\BibitemShut {NoStop}%
\bibitem [{\citenamefont {Gandikota}\ and\ \citenamefont
  {Schwarz}(2021)}]{Gandikota2021}%
  \BibitemOpen
  \bibfield  {author} {\bibinfo {author} {\bibfnamefont {M.~C.}\ \bibnamefont
  {Gandikota}}\ and\ \bibinfo {author} {\bibfnamefont {J.~M.}\ \bibnamefont
  {Schwarz}},\ }\bibfield  {title} {\bibinfo {title} {Buckling without bending
  morphogenesis: nonlinearities, spatial confinement, and branching
  hierarchies},\ }\href@noop {} {\bibfield  {journal} {\bibinfo  {journal} {New
  J. Phys.}\ }\textbf {\bibinfo {volume} {23}},\ \bibinfo {pages} {063060}
  (\bibinfo {year} {2021})}\BibitemShut {NoStop}%
\bibitem [{\citenamefont {{Borzou}}\ \emph {et~al.}(2021)\citenamefont
  {{Borzou}}, \citenamefont {{Patteson}},\ and\ \citenamefont
  {{Schwarz}}}]{Borzou2021}%
  \BibitemOpen
  \bibfield  {author} {\bibinfo {author} {\bibfnamefont {A.}~\bibnamefont
  {{Borzou}}}, \bibinfo {author} {\bibfnamefont {A.~E.}\ \bibnamefont
  {{Patteson}}},\ and\ \bibinfo {author} {\bibfnamefont {J.~M.}\ \bibnamefont
  {{Schwarz}}},\ }\bibfield  {title} {\bibinfo {title} {{A Data-Driven
  Statistical Description for the Hydrodynamics of Active Matter}},\
  }\href@noop {} {\bibfield  {journal} {\bibinfo  {journal} {arXiv e-prints}\
  ,\ \bibinfo {eid} {arXiv:2103.03461}} (\bibinfo {year} {2021})},\ \Eprint
  {https://arxiv.org/abs/2103.03461} {arXiv:2103.03461 [cond-mat.soft]}
  \BibitemShut {NoStop}%
\bibitem [{\citenamefont {Martin}\ and\ \citenamefont
  {Evans}(1975)}]{Martin1975}%
  \BibitemOpen
  \bibfield  {author} {\bibinfo {author} {\bibfnamefont {G.~R.}\ \bibnamefont
  {Martin}}\ and\ \bibinfo {author} {\bibfnamefont {M.~J.}\ \bibnamefont
  {Evans}},\ }\bibfield  {title} {\bibinfo {title} {Differentiation of clonal
  lines of teratocarcinoma cells: Formation of embryoid bodies in vitro},\
  }\href@noop {} {\bibfield  {journal} {\bibinfo  {journal} {Proc. Natl. Acad.
  Sci. USA}\ }\textbf {\bibinfo {volume} {72}},\ \bibinfo {pages} {1441}
  (\bibinfo {year} {1975})}\BibitemShut {NoStop}%
\bibitem [{\citenamefont {Murrell}\ \emph {et~al.}(2015)\citenamefont
  {Murrell}, \citenamefont {Oakes}, \citenamefont {Lenz},\ and\ \citenamefont
  {Gardel}}]{Murrell2015}%
  \BibitemOpen
  \bibfield  {author} {\bibinfo {author} {\bibfnamefont {M.}~\bibnamefont
  {Murrell}}, \bibinfo {author} {\bibfnamefont {P.}~\bibnamefont {Oakes}},
  \bibinfo {author} {\bibfnamefont {M.}~\bibnamefont {Lenz}},\ and\ \bibinfo
  {author} {\bibfnamefont {M.}~\bibnamefont {Gardel}},\ }\bibfield  {title}
  {\bibinfo {title} {Forcing cells into shapes: the mechanics of actomyosin
  contractility},\ }\href@noop {} {\bibfield  {journal} {\bibinfo  {journal}
  {Nat. Rev. Mol. Cell Biol.}\ }\textbf {\bibinfo {volume} {16}},\ \bibinfo
  {pages} {486} (\bibinfo {year} {2015})}\BibitemShut {NoStop}%
\bibitem [{\citenamefont {Humphrey}\ \emph {et~al.}(2014)\citenamefont
  {Humphrey}, \citenamefont {Dufresne},\ and\ \citenamefont
  {Schwartz}}]{Humphrey2014}%
  \BibitemOpen
  \bibfield  {author} {\bibinfo {author} {\bibfnamefont {J.~D.}\ \bibnamefont
  {Humphrey}}, \bibinfo {author} {\bibfnamefont {E.~R.}\ \bibnamefont
  {Dufresne}},\ and\ \bibinfo {author} {\bibfnamefont {M.~A.}\ \bibnamefont
  {Schwartz}},\ }\bibfield  {title} {\bibinfo {title} {Mechanotransduction and
  extracellular homeostasis},\ }\href@noop {} {\bibfield  {journal} {\bibinfo
  {journal} {Nat. Rev. Mol. Cell Biol.}\ }\textbf {\bibinfo {volume} {15}},\
  \bibinfo {pages} {802} (\bibinfo {year} {2014})}\BibitemShut {NoStop}%
\bibitem [{\citenamefont {Steinhardt}\ \emph {et~al.}(2005)\citenamefont
  {Steinhardt}, \citenamefont {Mukhanov}, \citenamefont {Mukhanov},
  \citenamefont {Linde}, \citenamefont {Press},\ and\ \citenamefont
  {Viatcheslav}}]{steinhardt2005physical}%
  \BibitemOpen
  \bibfield  {author} {\bibinfo {author} {\bibfnamefont {P.}~\bibnamefont
  {Steinhardt}}, \bibinfo {author} {\bibfnamefont {V.}~\bibnamefont
  {Mukhanov}}, \bibinfo {author} {\bibfnamefont {V.}~\bibnamefont {Mukhanov}},
  \bibinfo {author} {\bibfnamefont {A.}~\bibnamefont {Linde}}, \bibinfo
  {author} {\bibfnamefont {C.~U.}\ \bibnamefont {Press}},\ and\ \bibinfo
  {author} {\bibfnamefont {M.}~\bibnamefont {Viatcheslav}},\ }\href
  {https://books.google.com/books?id=1TXO7GmwZFgC} {\emph {\bibinfo {title}
  {Physical Foundations of Cosmology}}},\ Physical Foundations of Cosmology\
  (\bibinfo  {publisher} {Cambridge University Press},\ \bibinfo {year}
  {2005})\BibitemShut {NoStop}%
\bibitem [{\citenamefont {Binney}\ and\ \citenamefont
  {Tremaine}(2008)}]{binney2008galactic}%
  \BibitemOpen
  \bibfield  {author} {\bibinfo {author} {\bibfnamefont {J.}~\bibnamefont
  {Binney}}\ and\ \bibinfo {author} {\bibfnamefont {S.}~\bibnamefont
  {Tremaine}},\ }\href {https://books.google.com/books?id=qxWt20TH--cC} {\emph
  {\bibinfo {title} {Galactic Dynamics: Second Edition}}},\ Princeton Series in
  Astrophysics\ (\bibinfo  {publisher} {Princeton University Press},\ \bibinfo
  {year} {2008})\BibitemShut {NoStop}%
\bibitem [{\citenamefont {Honda}(1978)}]{Honda1978}%
  \BibitemOpen
  \bibfield  {author} {\bibinfo {author} {\bibfnamefont {H.}~\bibnamefont
  {Honda}},\ }\bibfield  {title} {\bibinfo {title} {Description of cellular
  patterns by dirichlet domains: the two-dimensional case.},\ }\href@noop {}
  {\bibfield  {journal} {\bibinfo  {journal} {J. Theor. Biol.}\ }\textbf
  {\bibinfo {volume} {72}},\ \bibinfo {pages} {523} (\bibinfo {year}
  {1978})}\BibitemShut {NoStop}%
\bibitem [{\citenamefont {Kaliman}\ \emph {et~al.}(2016)\citenamefont
  {Kaliman}, \citenamefont {Jayachandran}, \citenamefont {Rehfeldt},\ and\
  \citenamefont {Smith}}]{Kaliman2016}%
  \BibitemOpen
  \bibfield  {author} {\bibinfo {author} {\bibfnamefont {S.}~\bibnamefont
  {Kaliman}}, \bibinfo {author} {\bibfnamefont {C.}~\bibnamefont
  {Jayachandran}}, \bibinfo {author} {\bibfnamefont {F.}~\bibnamefont
  {Rehfeldt}},\ and\ \bibinfo {author} {\bibfnamefont {A.-S.}\ \bibnamefont
  {Smith}},\ }\bibfield  {title} {\bibinfo {title} {Limits of applicability of
  the voronoi tessellation determined by centers of cell nuclei to epithelium
  morphology},\ }\href@noop {} {\bibfield  {journal} {\bibinfo  {journal}
  {Frontiers in Physiology}\ }\textbf {\bibinfo {volume} {7}},\ \bibinfo
  {pages} {551} (\bibinfo {year} {2016})}\BibitemShut {NoStop}%
\bibitem [{\citenamefont {Versaevel}\ \emph {et~al.}(2012)\citenamefont
  {Versaevel}, \citenamefont {Grevesse},\ and\ \citenamefont
  {Gabriele}}]{Versaevel2012}%
  \BibitemOpen
  \bibfield  {author} {\bibinfo {author} {\bibfnamefont {M.}~\bibnamefont
  {Versaevel}}, \bibinfo {author} {\bibfnamefont {T.}~\bibnamefont
  {Grevesse}},\ and\ \bibinfo {author} {\bibfnamefont {S.}~\bibnamefont
  {Gabriele}},\ }\bibfield  {title} {\bibinfo {title} {Spatial coordination
  between cell and nuclear shape within micropatterned endothelial cells},\
  }\href@noop {} {\bibfield  {journal} {\bibinfo  {journal} {Nat. Commun.}\
  }\textbf {\bibinfo {volume} {3}},\ \bibinfo {pages} {671} (\bibinfo {year}
  {2012})}\BibitemShut {NoStop}%
\bibitem [{\citenamefont {Yevick}\ \emph {et~al.}(2019)\citenamefont {Yevick},
  \citenamefont {Miller}, \citenamefont {Dunkel},\ and\ \citenamefont
  {Martin}}]{Yevick2019}%
  \BibitemOpen
  \bibfield  {author} {\bibinfo {author} {\bibfnamefont {H.~G.}\ \bibnamefont
  {Yevick}}, \bibinfo {author} {\bibfnamefont {P.~W.}\ \bibnamefont {Miller}},
  \bibinfo {author} {\bibfnamefont {J.}~\bibnamefont {Dunkel}},\ and\ \bibinfo
  {author} {\bibfnamefont {A.~C.}\ \bibnamefont {Martin}},\ }\bibfield  {title}
  {\bibinfo {title} {Structural redundancy in supracellular actomyosin networks
  enables robust tissue folding},\ }\href@noop {} {\bibfield  {journal}
  {\bibinfo  {journal} {Developmental Cell}\ }\textbf {\bibinfo {volume}
  {50}},\ \bibinfo {pages} {586} (\bibinfo {year} {2019})}\BibitemShut
  {NoStop}%
\bibitem [{\citenamefont {Greiner}\ \emph {et~al.}(2013)\citenamefont
  {Greiner}, \citenamefont {Chen}, \citenamefont {Spatz},\ and\ \citenamefont
  {Kemkemer}}]{Greiner2013}%
  \BibitemOpen
  \bibfield  {author} {\bibinfo {author} {\bibfnamefont {A.~M.}\ \bibnamefont
  {Greiner}}, \bibinfo {author} {\bibfnamefont {H.}~\bibnamefont {Chen}},
  \bibinfo {author} {\bibfnamefont {J.~P.}\ \bibnamefont {Spatz}},\ and\
  \bibinfo {author} {\bibfnamefont {R.}~\bibnamefont {Kemkemer}},\ }\bibfield
  {title} {\bibinfo {title} {Cyclic tensile strain controls cell shape and
  directs active stress fiber formation and focal adhesion alignment in
  spreading cells},\ }\href@noop {} {\bibfield  {journal} {\bibinfo  {journal}
  {PLoS ONE}\ }\textbf {\bibinfo {volume} {8}},\ \bibinfo {pages} {e77328}
  (\bibinfo {year} {2013})}\BibitemShut {NoStop}%
\bibitem [{\citenamefont {Van~Oosten}\ \emph {et~al.}(2016)\citenamefont
  {Van~Oosten}, \citenamefont {Vahabi}, \citenamefont {Licup}, \citenamefont
  {Sharma}, \citenamefont {Galie}, \citenamefont {MacKintosh},\ and\
  \citenamefont {Janmey}}]{VanOosten2016}%
  \BibitemOpen
  \bibfield  {author} {\bibinfo {author} {\bibfnamefont {A.~S.~G.}\
  \bibnamefont {Van~Oosten}}, \bibinfo {author} {\bibfnamefont
  {M.}~\bibnamefont {Vahabi}}, \bibinfo {author} {\bibfnamefont {A.~J.}\
  \bibnamefont {Licup}}, \bibinfo {author} {\bibfnamefont {A.}~\bibnamefont
  {Sharma}}, \bibinfo {author} {\bibfnamefont {P.~A.}\ \bibnamefont {Galie}},
  \bibinfo {author} {\bibfnamefont {F.~C.}\ \bibnamefont {MacKintosh}},\ and\
  \bibinfo {author} {\bibfnamefont {P.~A.}\ \bibnamefont {Janmey}},\ }\bibfield
   {title} {\bibinfo {title} {Uncoupling shear and uniaxial elastic moduli of
  semiflexible biopolymer networks: compression-softening and
  stretch-stiffening},\ }\href@noop {} {\bibfield  {journal} {\bibinfo
  {journal} {Sci. Reps.}\ }\textbf {\bibinfo {volume} {6}},\ \bibinfo {pages}
  {19270} (\bibinfo {year} {2016})}\BibitemShut {NoStop}%
\bibitem [{\citenamefont {Larsell}(1967)}]{Larsell}%
  \BibitemOpen
  \bibfield  {author} {\bibinfo {author} {\bibfnamefont {O.}~\bibnamefont
  {Larsell}},\ }\href@noop {} {\emph {\bibinfo {title} {The Comparative Anatomy
  and Histology of the Cerebellum}}}\ (\bibinfo  {publisher} {University of
  Minnesota Press},\ \bibinfo {year} {1967})\BibitemShut {NoStop}%
\bibitem [{\citenamefont {Sudarov}\ and\ \citenamefont
  {Joyner}(2007)}]{Sudarov2007}%
  \BibitemOpen
  \bibfield  {author} {\bibinfo {author} {\bibfnamefont {A.}~\bibnamefont
  {Sudarov}}\ and\ \bibinfo {author} {\bibfnamefont {A.~L.}\ \bibnamefont
  {Joyner}},\ }\bibfield  {title} {\bibinfo {title} {Cerebellum morphogenesis:
  the foliation pattern is orchestrated by multi-cellular anchoring centers},\
  }\href@noop {} {\bibfield  {journal} {\bibinfo  {journal} {Neural
  Development}\ }\textbf {\bibinfo {volume} {2}},\ \bibinfo {pages} {26}
  (\bibinfo {year} {2007})}\BibitemShut {NoStop}%
\bibitem [{\citenamefont {Mongera}\ and\ \citenamefont
  {et~al.}(2018)}]{mongera_2018}%
  \BibitemOpen
  \bibfield  {author} {\bibinfo {author} {\bibfnamefont {A.}~\bibnamefont
  {Mongera}}\ and\ \bibinfo {author} {\bibnamefont {et~al.}},\ }\bibfield
  {title} {\bibinfo {title} {A fluid-to-solid jamming transition underlies
  vertebrate body axis elongation},\ }\href@noop {} {\bibfield  {journal}
  {\bibinfo  {journal} {Nature}\ }\textbf {\bibinfo {volume} {561}},\ \bibinfo
  {pages} {401} (\bibinfo {year} {2018})}\BibitemShut {NoStop}%
\bibitem [{\citenamefont {Jain}\ and\ \citenamefont {et~al}(2020)}]{jain_2020}%
  \BibitemOpen
  \bibfield  {author} {\bibinfo {author} {\bibfnamefont {A.}~\bibnamefont
  {Jain}}\ and\ \bibinfo {author} {\bibnamefont {et~al}},\ }\bibfield  {title}
  {\bibinfo {title} {Regionalized tissue fluidization is required for
  epithelial gap closure during insect gastrulation},\ }\href@noop {}
  {\bibfield  {journal} {\bibinfo  {journal} {Nature Communications}\ }\textbf
  {\bibinfo {volume} {11}},\ \bibinfo {pages} {5604} (\bibinfo {year}
  {2020})}\BibitemShut {NoStop}%
\bibitem [{\citenamefont {Kim}\ \emph {et~al.}(2021)\citenamefont {Kim},
  \citenamefont {Pochitaloff}, \citenamefont {Stooke-Vaughan},\ and\
  \citenamefont {Campas}}]{Kim2021}%
  \BibitemOpen
  \bibfield  {author} {\bibinfo {author} {\bibfnamefont {S.}~\bibnamefont
  {Kim}}, \bibinfo {author} {\bibfnamefont {M.}~\bibnamefont {Pochitaloff}},
  \bibinfo {author} {\bibfnamefont {G.~A.}\ \bibnamefont {Stooke-Vaughan}},\
  and\ \bibinfo {author} {\bibfnamefont {O.}~\bibnamefont {Campas}},\
  }\bibfield  {title} {\bibinfo {title} {Embryonic tissues as active foams},\
  }\href@noop {} {\bibfield  {journal} {\bibinfo  {journal} {Nature Phys.}\
  }\textbf {\bibinfo {volume} {17}},\ \bibinfo {pages} {859} (\bibinfo {year}
  {2021})}\BibitemShut {NoStop}%
\bibitem [{\citenamefont {Brandt}\ \emph {et~al.}(2005)\citenamefont {Brandt},
  \citenamefont {Rohlfing}, \citenamefont {Rybak}, \citenamefont {Krofczik},
  \citenamefont {Maye}, \citenamefont {Westerhoff}, \citenamefont {Hege},\ and\
  \citenamefont {Menzel}}]{Brandt2005}%
  \BibitemOpen
  \bibfield  {author} {\bibinfo {author} {\bibfnamefont {R.}~\bibnamefont
  {Brandt}}, \bibinfo {author} {\bibfnamefont {R.}~\bibnamefont {Rohlfing}},
  \bibinfo {author} {\bibfnamefont {J.}~\bibnamefont {Rybak}}, \bibinfo
  {author} {\bibfnamefont {S.}~\bibnamefont {Krofczik}}, \bibinfo {author}
  {\bibfnamefont {A.}~\bibnamefont {Maye}}, \bibinfo {author} {\bibfnamefont
  {M.}~\bibnamefont {Westerhoff}}, \bibinfo {author} {\bibfnamefont {H.-C.}\
  \bibnamefont {Hege}},\ and\ \bibinfo {author} {\bibfnamefont
  {R.}~\bibnamefont {Menzel}},\ }\bibfield  {title} {\bibinfo {title}
  {Three-dimensional average-shape atlas of the honeybee brain and its
  applications},\ }\href@noop {} {\bibfield  {journal} {\bibinfo  {journal} {J.
  Comp. Neurol.}\ }\textbf {\bibinfo {volume} {492}},\ \bibinfo {pages} {1}
  (\bibinfo {year} {2005})}\BibitemShut {NoStop}%
\bibitem [{\citenamefont {{Guth}}\ and\ \citenamefont
  {{Pi}}(1982)}]{1982PhRvL..49.1110G}%
  \BibitemOpen
  \bibfield  {author} {\bibinfo {author} {\bibfnamefont {A.~H.}\ \bibnamefont
  {{Guth}}}\ and\ \bibinfo {author} {\bibfnamefont {S.~Y.}\ \bibnamefont
  {{Pi}}},\ }\bibfield  {title} {\bibinfo {title} {{Fluctuations in the New
  Inflationary Universe}},\ }\href
  {https://doi.org/10.1103/PhysRevLett.49.1110} {\bibfield  {journal} {\bibinfo
   {journal} {\prl}\ }\textbf {\bibinfo {volume} {49}},\ \bibinfo {pages}
  {1110} (\bibinfo {year} {1982})}\BibitemShut {NoStop}%
\bibitem [{\citenamefont {{Starobinsky}}(1982)}]{1982PhLB..117..175S}%
  \BibitemOpen
  \bibfield  {author} {\bibinfo {author} {\bibfnamefont {A.~A.}\ \bibnamefont
  {{Starobinsky}}},\ }\bibfield  {title} {\bibinfo {title} {{Dynamics of phase
  transition in the new inflationary universe scenario and generation of
  perturbations}},\ }\href {https://doi.org/10.1016/0370-2693(82)90541-X}
  {\bibfield  {journal} {\bibinfo  {journal} {Physics Letters B}\ }\textbf
  {\bibinfo {volume} {117}},\ \bibinfo {pages} {175} (\bibinfo {year}
  {1982})}\BibitemShut {NoStop}%
\bibitem [{\citenamefont {{Hawking}}(1982)}]{1982PhLB..115..295H}%
  \BibitemOpen
  \bibfield  {author} {\bibinfo {author} {\bibfnamefont {S.~W.}\ \bibnamefont
  {{Hawking}}},\ }\bibfield  {title} {\bibinfo {title} {{The development of
  irregularities in a single bubble inflationary universe}},\ }\href
  {https://doi.org/10.1016/0370-2693(82)90373-2} {\bibfield  {journal}
  {\bibinfo  {journal} {Physics Letters B}\ }\textbf {\bibinfo {volume}
  {115}},\ \bibinfo {pages} {295} (\bibinfo {year} {1982})}\BibitemShut
  {NoStop}%
\end{thebibliography}%



\setcounter{equation}{0}
\setcounter{figure}{0}
\setcounter{table}{0}
\setcounter{page}{1}
\makeatletter
\renewcommand{\theequation}{A\arabic{equation}}
\renewcommand{\thefigure}{A\arabic{figure}}
\renewcommand{\bibnumfmt}[1]{[A#1]}
\appendix

\section{Stage I analysis} 
\lb{App:Stage1}
The conventional approach to solving
Eq.~\eqref{Eq:StochasticDynamicEq} is to compute its first two
moments with respect to velocities leading to two differential equations for the
number density $\rho$, the bulk velocity $\bar{v}^j$, and
$\overline{vv}^{ij}$ each defined as 
\bqn
\lb{Eq:rho_vbar_vvbar}
&&\rho \equiv \int dv \, f,\nb\\
&&\bar{v}^j \equiv \frac{1}{\rho}\int dv \, v^j\, f,\nb\\
&&\overline{vv}^{ij} \equiv \frac{1}{\rho}\int dv \, v^iv^j\, f  
\eqn
such that 
\bqn
\lb{Eq:MomentEqsSM}
&&\frac{\partial \rho}{\partial t} + \partial_i \left( \rho \bar{v}^i \right) = \int dv \, C,\nb\\
&&\frac{\partial}{\partial t} \left( \rho \bar{v}^j \right) + \partial_i \left( \rho \overline{vv}^{ij}\right)
+ \rho g^j = \int dv \, C \, v^j.
\eqn
However, a common problem with this approach is that there are three unknown
variables, but
only two differential equations in Eq.~\eqref{Eq:MomentEqsSM}.  One can
find another differential equation for $\overline{vv}^{ij}$, which
would depend on higher moments of $f$.  A common remedy is to assume a
relationship between the number density and the pressure of the
system.  To do so, we define the stress tensor as 
\bqn
\sigma^{2ij}  \equiv \overline{vv}^{ij} - \bar{v}^i\bar{v}^j.
\eqn
Since the system looks the same at different angles at least at the
beginning stages,  we assume that the stress tensor is isotropic and 
write $\sigma^{2ij} = \sigma^2 \delta^{ij}$.  We then use data from
the experiment movies reported in Ref.~\cite{Karzbrun2018} and applied
our data-driven method presented in Ref.~\cite{Borzou2021} to
find that the pressure of the cell nuclei
linearly depends on the number density with a proportionality
coefficient of $\sigma^2=0.1$ (see Appendix 
Fig.~\ref{Fig:P_rho_fit}).  With this finding, the final form of the evolution equation set reads
\bqn
&&\partial_t \rho + \partial_i \left( \rho \bar{v}_i \right) = C_0,\nb\\
&&\partial_t \bar{v}_j + \sigma^2 \partial_j \rho + \bar{v}_i \partial_i \bar{v}_j + g_j = \frac{1}{\rho} \left(C_j - \bar{v}_j C_0 \right),\lb{Eq:MomentEqsSM2}
\eqn
where
$C_0 \equiv \int dv \, C$,  which account for cell division and the
noise,  and $C_j \equiv  \int dv \, C \, v_j$,  which accounts for
dissipation and noise. More details regarding $C_0$ and $C_j$ are
discussed in the main section of the manuscript.  The evolution
equations are then solved in the linear regime after which time the
final condition for the linear regime is then used as an initial
condition for the nonlinear regime. 

\begin{figure}[h]
\centering
\includegraphics[width=0.5\textwidth]{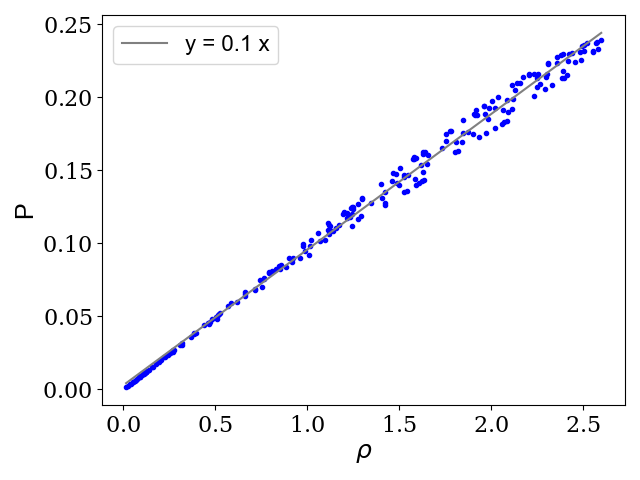}
\caption{Relationship between the pressure and cell nuclei density
  extracted from Supplementary Movie II from Ref.~\cite{Karzbrun2018}
  using the data-driven technique presented in Ref.~\cite{Borzou2021}.}
\label{Fig:P_rho_fit}
\end{figure}

\subsection{Linear evolution}
\lb{Sec:LinearStochDynamic}

In the linear regime,  we assume that the number density is
initially homogeneous with some small fluctuations, or  
\bqn
\rho \equiv \rho_0 + \delta \rho,
\eqn
with $\delta \rho \ll \rho_0$.  Inserting this equation into
Eqs.~\eqref{Eq:MomentEqsSM2} and neglecting higher-order terms,
they read
\bqn
\lb{Eq:MomentEqs_linearSM}
&&\frac{\partial}{\partial t}\delta \rho + \rho_0 \partial_i\bar{v}_i = C_{0_{\text{noise}}},\nb\\
&&\frac{\partial}{\partial t}\bar{v}_j + \sigma^2 \partial_j \delta \rho +  \left(g_j- C_j/\rho_0 \right)= 0,
\eqn
where we have neglected $\bar{v}^2$ and $C_{0_{\text{noise}}}\bar{v}$
since they are of the order of ${\cal{O}}(\delta \rho^2)$. Moreover,
we have assumed that the bulk velocity $\bar{v}_i$,  the damping
contribution to $C_j$,  and the coefficient of the pressure $\sigma^2$ are all
small and of the order of ${\cal{O}}(\delta \rho)$. The effects of
damping are discussed in the next subsection.

To work towards a solution, we Fourier transform Eq.~\eqref{Eq:MomentEqs_linearSM} to arrive at 
\bqn
\lb{Eq:MomentEqs_Fourier_v1SM}
&&\partial_t \tilde{\delta} + ik_i\rho_0 \tilde{v}_i=\tilde{C}_{0_{\text{noise}}},\nb\\
&&\partial_t \tilde{v}_j + \sigma^2 i k_j \tilde{\delta} + \tilde{g}_j = \tilde{C}_{j_{\text{noise}}}/\rho_0,
\eqn
where $\tilde{\delta}$, $\tilde{v}$ ,  and $\tilde{g}_j$ refer to the $k$ mode of the Fourier transformations of $\delta\rho$,  
$\bar{v}$,  and $g_j- C_{j_{\text{no-noise}}}/\rho_0$ respectively.
Also,  $\tilde{C}_{0_{\text{noise}}}$ and
$\tilde{C}_{j_{\text{noise}}}$ denote the Fourier transformation of the
respective noise contributions and obey $\langle
\tilde{C}_{0_{\text{noise}}} \rangle = \langle
\tilde{C}_{j_{\text{noise}}} \rangle =0$, along with 
\bqn
&&\langle \tilde{C}_{0_{\text{noise}}} \tilde{C}_{0_{\text{noise}}} \rangle = \left(2\pi\right)^3\theta \delta(t-t')\delta^3(\vec{k}+\vec{k}').\nb\\
&&\langle \tilde{C}_{i_{\text{noise}}}(t,\vec{k}) \tilde{C}_{j_{\text{noise}}}(t',\vec{k}')\rangle = \left(2\pi\right)^3\gamma\times \nb\\
&& \delta_{ij} \delta(t-t')\delta^3(\vec{k}+\vec{k}'). 
\eqn
With this form for the noise, the variables are functions of the same mode $k$,  i.e.  mode $k'$ and $k$ are independent. To proceed, we assume that 
\bqn
\lb{Eq:FourierForceSM}
\tilde{g}_j  = - i k_j {\cal{L}}^{-1} \tilde{\delta}.
\eqn
As discussed in the main text, we choose a specific form
$\cal{L}^{-1}$ and continue with the
formal solution for the evolution of initial overdensities. We define $\tilde{X}\equiv i k_i \tilde{v}_i$ and multiply the second line of Eq.~\eqref{Eq:MomentEqs_Fourier_v1SM} by $i k_j$ to write the set of equations as
\bqn
\lb{Eq:MomentEqs_Fourier_v3SM}
&&\partial_t \tilde{\delta} + \rho_0 \tilde{X}=\tilde{C}_{0_{\text{noise}}},\nb\\
&&\partial_t \tilde{X} + k^2\left({\cal{L}}^{-1}-\sigma^2 \right)\tilde{\delta} = ik_j\tilde{C}_{j_{\text{noise}}}/\rho_0,
\eqn
which can be written in the following matrix form
\bqn
\lb{Eq:MomentEqs_Fourier_v2SM}
\partial_t 
\begin{pmatrix}
\tilde{\delta}\\
\tilde{X}
\end{pmatrix}
+
M
\cdot
\begin{pmatrix}
\tilde{\delta}\\
\tilde{X}
\end{pmatrix}
=
\begin{pmatrix}
~~\\
\text{noise}\\
~~\\
\end{pmatrix}, 
\eqn
where 
\bqn
\hat{M} = 
\begin{pmatrix}
0 & \rho_0\\
k^2\left({\cal{L}}^{-1}-\sigma^2\right) & 0
\end{pmatrix}.
\eqn

To solve Eq.~\eqref{Eq:MomentEqs_Fourier_v2SM},  we decouple the
two equations by diagonalizing matrix $\hat{M}$ with matrix $\hat{U}^{-1}$  and rewrite it in the following form
\bqn
\partial_t \tilde{Y} + \hat{M}_d \cdot \tilde{Y} = \begin{pmatrix}
~~\\
\text{noise}'\\
~~\\
\end{pmatrix} ,
\eqn
where 
\bqn
&&\tilde{Y} = U^{-1}\cdot 
\begin{pmatrix}
\tilde{\delta}\\
\tilde{X}
\end{pmatrix},\nb\\
&&
M_d = U^{-1}\cdot  M \cdot U.
\eqn
The prime on the noise means it is transformed by the $U^{-1}$ matrix
as well. 
We define $U$ such that $M_d$ is a diagonal matrix.  Therefore,  the solution to each component of $Y$ reads
\bqn
\tilde{Y}_i &=& \tilde{Y}_i(t=0)\exp\left(-M_{d_{ii}}t\right)\nb\\
&+&\text{noise}'_i\left(1 - \exp\left(-M_{d_{ii}}t\right) \right).
\eqn
This equation indicates that the noise is irrelevant for the modes
that grow over time and become seeds for the large-scale structures
in the non-linear regime.  The reason is that the growing modes are
those in which $M_{d_{ii}}$ is negative and for these modes the noise
term in right hand side of the equation becomes rapidly neligible. It should be noted that the linear regime is only valid
until $\tilde{Y}<1$ and the noise term will not appear with a negative
coefficient at $t\gg 0$.  Interestingly,  the same exponentially
growing modes are the seeds to the galaxies in the cosmos.  Finally, the solutions to the variables of interest are
\bqn
\begin{pmatrix}
\tilde{\delta}(t,\vec{k})\\
~\\
\tilde{X}(t,\vec{k})
\end{pmatrix}
=
U
\cdot 
\begin{pmatrix}
\tilde{Y}_1\\
~~\\
\tilde{Y}_2
\end{pmatrix}.
\eqn

At this point, we have solved the differential equations in Fourier
space and now convert back to configuration space where the solution reads
\bqn
\lb{Eq:deltaRhoSolSM}
\delta \rho(t,\vec{x})=\int d^3k\, e^{i\vec{k}\cdot\vec{k}} \, \tilde{\delta}(t,\vec{k}),
\eqn
where 
\bqn
\tilde{\delta}(t,\vec{k}) = \tilde{\delta}(t=0,\vec{k}) \cosh\left(\sqrt{\rho_0 ( {\cal{L}}^{-1} -\sigma^2) } ~k t\right).
\eqn
Since ${\cal{L}}^{-1}$ is also a function of $k$,  for some modes the
square root takes imaginary values, leading to oscillations, which we
do not focus on here. This model can readily be tested against observations. 
An interesting set of observable quantities in the linear regime is the time-evolved correlations between the overdensities $\langle \delta\rho(t, x_1)\cdots \delta\rho(t,x_n)\rangle$.
Therefore,  one can determine the form of ${\cal{L}}^{-1}$ using a few
of these correlation functions and further interrogate the model using the rest of the correlation functions.

\begin{figure*}
\includegraphics[width=0.3\textwidth]{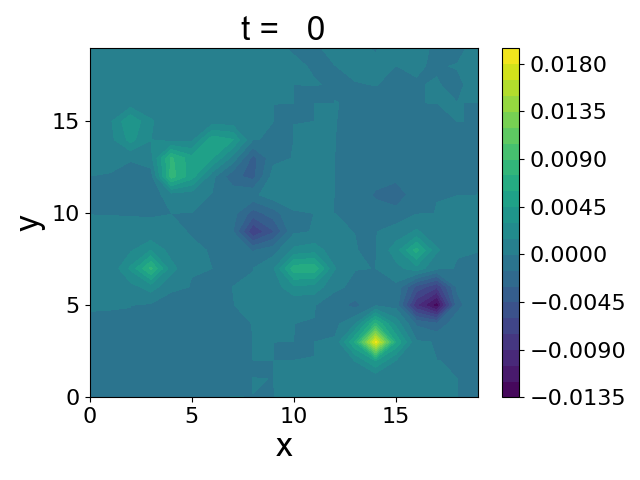}
\includegraphics[width=0.3\textwidth]{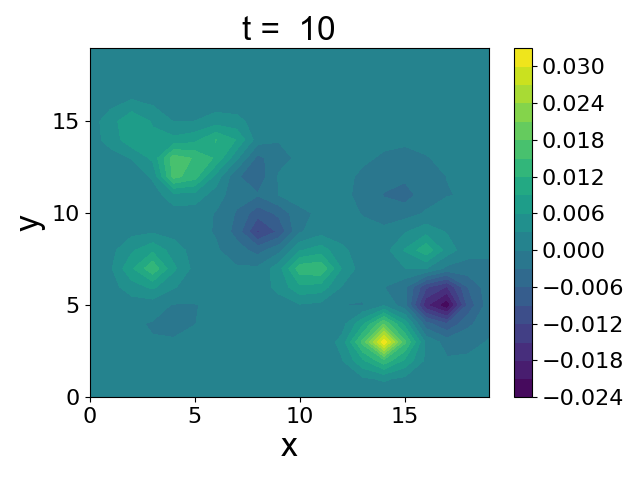}
\includegraphics[width=0.3\textwidth]{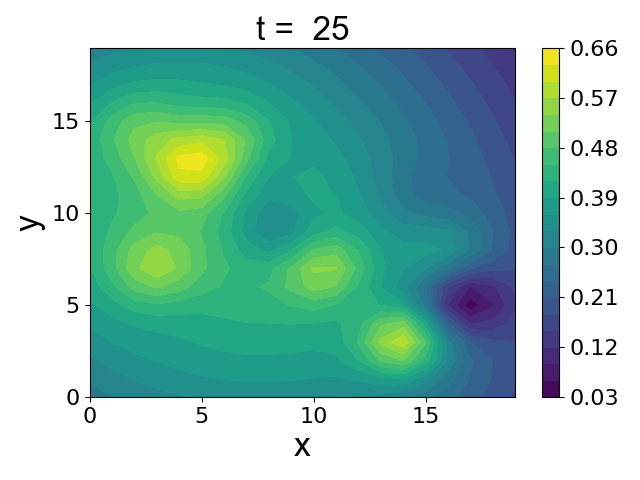}
\includegraphics[width=0.3\textwidth]{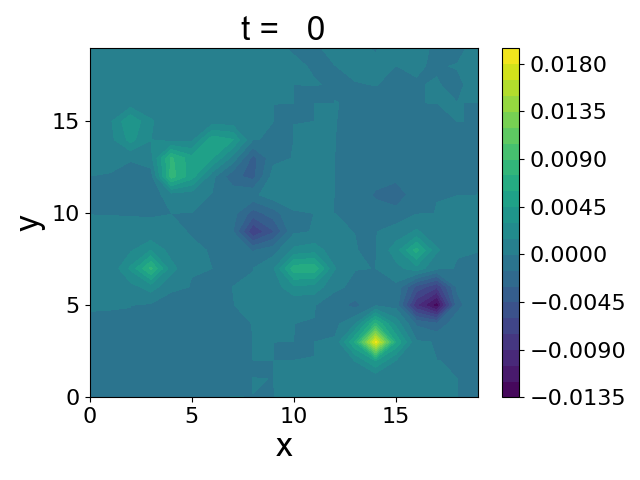}
\includegraphics[width=0.3\textwidth]{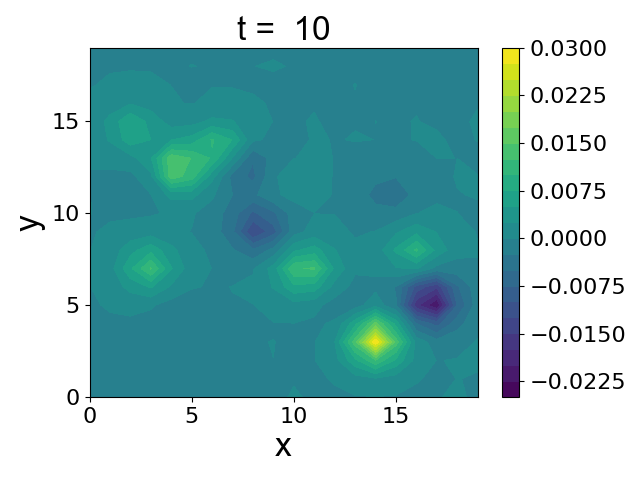}
\includegraphics[width=0.3\textwidth]{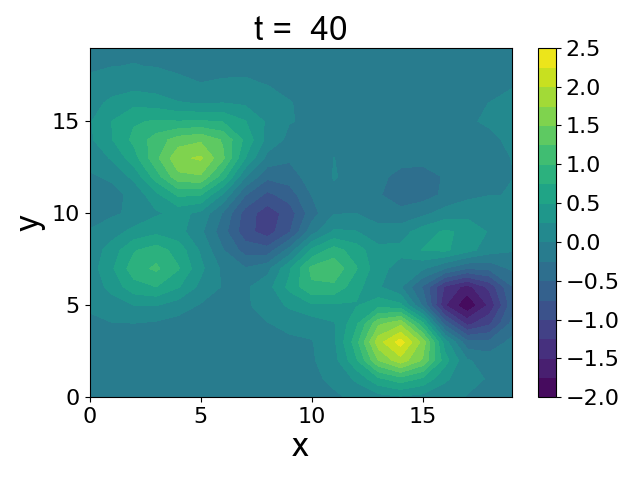}
\caption{Top row: The time evolution for $\delta\rho$ as a function of
  positions $x$ and $y$ for ${\cal{L}}^{-1} =
  \frac{1}{k^2+0.1^2}+\frac{0.1}{k^4+0.1^4}$. Bottom row: Same as top
  row but with ${\cal{L}}^{-1} =
  \frac{1}{k^2+0.1^2}$, just as in the Fig. 2 in the main text.  
\lb{Fig:EvolutionOfRho_with2nd}}
\end{figure*}

\subsubsection{The Effects of Dissipation}
Here we study how a force proportional to the negative of velocity, to
model dissipation, affects the structure formation in the linear regime.  Therefore,  we replace $C_j \rightarrow C_j - a_2 \rho_0 \bar{v}_j$

\bqn
\lb{Eq:MomentEqs_linear}
&&\frac{\partial}{\partial t}\delta \rho + \rho_0 \partial_i\bar{v}_i = 0,\nb\\
&&\frac{\partial}{\partial t}\bar{v}_j + \sigma^2 \partial_j \delta \rho +  \left(g_j- C_j/\rho_0 \right) + a_2 \bar{v}_j = 0,
\eqn
which in momentum space reads
\bqn
\lb{Eq:MomentEqs_Fourier_v3}
&&\partial_t \tilde{\delta} + \rho_0 \tilde{X}=0,\nb\\
&&\partial_t \tilde{X} + k^2\left({\cal{L}}^{-1}-\sigma^2 \right)\tilde{\delta} + a_2 \tilde{X}= 0.
\eqn
This coupled system of equations can then be written in the following matrix form
\bqn
\lb{Eq:MomentEqs_Fourier_v2}
\partial_t 
\begin{pmatrix}
\tilde{\delta}\\
\tilde{X}
\end{pmatrix}
+
M
\cdot
\begin{pmatrix}
\tilde{\delta}\\
\tilde{X}
\end{pmatrix}
=
0,
\eqn
where 
\bqn
M = 
\begin{pmatrix}
0 & \rho_0\\
k^2\left({\cal{L}}^{-1}-\sigma^2\right) & a_2
\end{pmatrix}.
\eqn
Then  
\bqn
\tilde{\delta}(t,\vec{k}) = \tilde{\delta}(t=0,\vec{k}) 
e^{-\frac{a_2 t}{2}} \left( A + B \right),
\eqn
with 
\bqn
\nb\\
&&A \equiv 
\cosh \left(\frac{1}{2} t \sqrt{4 k^2 \rho_0 ({\cal{L}}^{-1}-\sigma^2)+a_2^2}\right),\nb\\
&&B \equiv
\frac{a_2 \sinh \left(\frac{1}{2} t \sqrt{4 k^2 \rho_0 ({\cal{L}}^{-1}-2
   a)+a_2^2}\right)}{\sqrt{4 k^2 \rho_0 ({\cal{L}}^{-1}-\sigma^2)+a_2^2}}
.
\eqn

Fourier transforming back to configuration space,  we integrate over $k$-modes from zero to infinity.  For some large $k$,   $4 k^2 \rho_0 ({\cal{L}}^{-1}-2
   a) \gg a_2^2$ at short times, 
\bqn
\tilde{\delta}(t,\vec{k}) \simeq \tilde{\delta}(t=0,\vec{k}) 
\exp\left(t k \sqrt{\rho_0 ({\cal{L}}^{-1}-\sigma^2)}\right). 
\eqn
On the other hand,  for small $k$-modes where $4 k^2 \rho_0
({\cal{L}}^{-1}-2a) \ll a_2^2$,  the over-densities will dissolve
after some time since 
\bqn
\tilde{\delta}(t,\vec{k}) \propto \tilde{\delta}(t=0,\vec{k}) 
\exp\left(-\frac{a_2 t}{2}\right). 
\eqn
In sum,  dissipation effects lead to removal of small $k$-modes, while
larger $k$-modes still grow over time and create seeds for the non-linear regime.   Observation of the smallest scales that form in the experiment can be used to set the dissipation effects using data.

\subsection{Non-linear evolution}
\lb{Sec:2DEvol}
We now study the non-linear regime of the large-scale structure
formation. We reset $t=0$ to the beginning of the non-linear regime.  At the end of the linear era we have
\bqn
\lb{Eq:InitialCondSM}
&&\rho \simeq \rho_0,\nb\\
&&\partial_t \rho \simeq 0,\nb\\
&&\bar{v}_r = - v_0,
\eqn
where in the first line we have used $\delta \rho \ll \rho_0$.  Also,
$v_0>0$ is the magnitude of the bulk velocity at the beginning of the
non-linear regime.  Since the cortex-core structures in the brain
organoid are nearly spherically symmetric, we write the closed system of differential equations as 
\bqn
\lb{Eq:MomentEqs_finalSM}
&&\frac{\partial \rho}{\partial t}+ \frac{2}{r}\rho\bar{v}_r + \partial_r \left( \rho \bar{v}_r \right) = C_0,\nb\\
&&\frac{\partial}{\partial t}  \bar{v}_r + \sigma^2 \partial_r \rho +\bar{v}_r \partial_r\bar{v}_r
+  g_r =\frac{1}{\rho} C_r ,
\eqn
where we have used the isotropic assumption to infer that in the spherical coordinate system $\vec{\bar{v}}=(\bar{v}_r,  0)$. 
We use the first differential equation in the exact evolution set in Eq.~\eqref{Eq:MomentEqs_finalSM} to eliminate the bulk velocity in terms of the number density 
\bqn
\lb{Eq:v_tr}
\bar{v}_r(t,r) = -\frac{1}{r^2\rho(t,r)}\int dr\,r^2 \partial_t \rho(t,r) + \frac{A(t)}{r^2\rho(t,r)}, 
\eqn
where $A(t)$ is an arbitrary function of time.  
Since the values of the number density and the bulk velocity at the
end of the linear regime serve as the initial conditions for the
non-linear regime,  the rest of the initial conditions are given in
Eq.~\eqref{Eq:InitialCondSM}.  Moreover, at the end of the
cortex-core formation, the system satisfies the conditions
\bqn
&&\rho \rightarrow F(r),\nb\\
&&A \rightarrow 0,\nb\\
&&\bar{v}_r \rightarrow  0,
\eqn
where $F(r)$ is the form of the cell nuclei number density that can be
observed at around Day 3 of the experiment~\cite{Karzbrun2018} and fit
it to a polynomial function.  The fit is plotted in Fig.~\ref{Fig:Fr}. 

\begin{figure}

\includegraphics[width=0.5\textwidth]{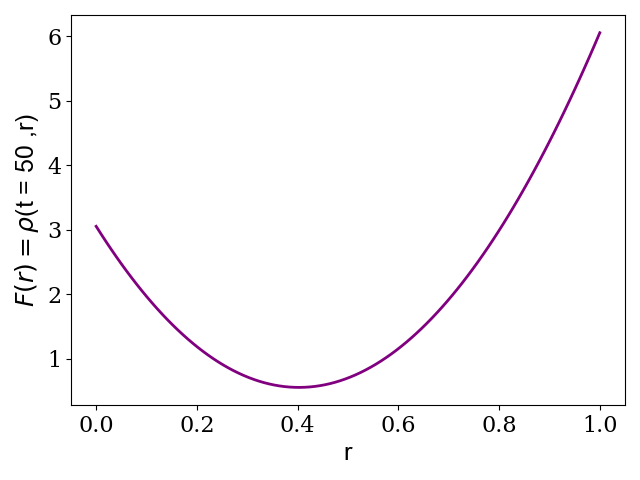}
\caption{The cell nuclei density
 $F(r)$ on Day 3 observed in the experiments in Ref~\cite{Karzbrun2018}.  }
\lb{Fig:Fr}
\end{figure}

Let us address the number of unknown variables and the means of determining them.   Unlike in cosmology,  the forces in the brain organoid system may
evolve in time due to restructuring of the cellular cytoskeleton in
response to the extracellular environment, for instance.  In the
linear regime, we assumed a minimal form for the interaction between similar
cell nuclei. As the brain organoid system evolves, the forces 
evolve. Hence,  $g_r(t,r)$ should be considered unknown.  In addition,
the damping effects in $C_r(t,r)$  and $A(t)$ are also unknown.  The
number density $\rho(t,r)$ is another unknown variable.  One of these
unknowns can be found by solving the remaining equation in
Eq.~\eqref{Eq:MomentEqs_finalSM}.  The rest should be found either
through data, and whenever data is not available, by assumption or
argument.  Accordingly, we construct an analytic form for $\rho(t,r)$ and
$A(t)$ using observations,  and solve the second line in
Eq.~\eqref{Eq:MomentEqs_finalSM} to determine the time evolution for the
net force on the cell nuclei,  i.e.
$g_r(t,r)-C_r(t,r)/\rho(t,r)$. We assume the number density to represented by the following profile
\bqn
\lb{Eq:rho_trSM}
\rho(t,r) = \rho_0 e^{-t/ \tau} + F(r) \left(1 - e^{t / \tau}\right),
\eqn
where $\tau$ is the half-life of the experiment equivalent to 1.5 days.  Moreover, given the initial and final conditions, we
assume that $A(t)=0$. We can now determine the time evolution of $\bar{v}_r(t,r)$  and $g_r(t,r)-C_r(t,r)/\rho(t,r)$.  The results are shown in Fig.~\ref{Fig:Fr}. 

Now that we have computed the net force on any given cell nuclei, we
ask what can we infer about nuclei shape. Assuming that the contractile
forces are given by Eq.~\eqref{Eq:FourierForceSM} throughout the
nonlinear regime, the net force of such interaction on a cell nuclei at position $\vec{x}$ is given by
\bqn
\lb{Eq:NetContracForceSM}
\vec{f}_{\text{cell-cell}}(\vec{x}) = i \int \frac{d^3k}{(2\pi)^3}\, \vec{k}\, {\cal{L}}^{-1} \int d^3x' \rho(\vec{x}')\,
e^{i\vec{k}\cdot (\vec{x}-\vec{x}')}.
\eqn
The remaining contribution of net force on a cell nuclei is due to
emergent interactions, either with other cells or with the
extracellular environment. To obtain Fig. 3b, we insert $F(r)$ into
the above equation, include first term of ${\cal{L}}^{-1}$ in Eq.~\eqref{Eq:Linv} ,
set $b=0.1$,  and calculate the integral in
Eq.~\ref{Eq:NetContracForceSM}.  The result is shown in Fig. 3b.  The
difference between this latter force and the net force in
Fig. 3b is the emergent force.

\end{document}